\documentclass[reprint,
superscriptaddress, 
 amsmath,amssymb,
 aps,
 pra,
onecolumn,
showkeys
]{revtex4-1}

\usepackage{color}
\usepackage{rotating}
\usepackage{graphicx}
\usepackage{dcolumn}
\usepackage{bm}
\usepackage{xcolor}
\usepackage{comment}
\usepackage[switch]{lineno} 

\begin{document}


\title{Six decades of the FitzHugh-Nagumo model:\\ A guide through its spatio-temporal dynamics and influence across disciplines }

\author{Daniel Cebri\'an-Lacasa}
 \affiliation{%
 Laboratory of Dynamics in Biological Systems, KU Leuven, Department of Cellular and Molecular Medicine, University of Leuven, B-3000 Leuven, Belgium.
}%
\affiliation{Centre for Engineering Biology, University of Edinburgh, Edinburgh EH9 3BF, United Kingdom.}

\author{Pedro Parra-Rivas}
 \affiliation{%
 Laboratory of Dynamics in Biological Systems, KU Leuven, Department of Cellular and Molecular Medicine, University of Leuven, B-3000 Leuven, Belgium.
}%
\affiliation{
Dipartimento di Ingegneria dell’Informazione, Elettronica e Telecomunicazioni, Sapienza Universitá di Roma, 00184 Rome, Italy.
}%

\author{Daniel Ruiz-Reyn\'es}
\affiliation{%
 Laboratory of Dynamics in Biological Systems, KU Leuven, Department of Cellular and Molecular Medicine, University of Leuven, B-3000 Leuven, Belgium.
}%
\affiliation{
IFISC (CSIC-UIB). Instituto de Física Interdisciplinar y Sistemas Complejos, E-07122 Palma de Mallorca, Spain
}

\author{Lendert Gelens}
\affiliation{%
 Laboratory of Dynamics in Biological Systems, KU Leuven, Department of Cellular and Molecular Medicine, University of Leuven, B-3000 Leuven, Belgium.
}%
\email{lendert.gelens@kuleuven.be}
 
\date{\today}
    
\begin{abstract}
    The FitzHugh-Nagumo equation, originally conceived in neuroscience during the 1960s, became a key model providing a simplified view of excitable neuron cell behavior. Its applicability, however, extends beyond neuroscience into fields like cardiac physiology, cell division, population dynamics, electronics, and other natural phenomena. In this review spanning six decades of research, we discuss the diverse spatio-temporal dynamical behaviors described by the FitzHugh-Nagumo equation. 
    These include dynamics like bistability, oscillations, and excitability, but it also addresses more complex phenomena such as traveling waves and extended patterns in coupled systems. The review serves as a guide for modelers aiming to utilize the strengths of the FitzHugh-Nagumo model to capture generic dynamical behavior. It not only catalogs known dynamical states and bifurcations, but also extends previous studies by providing stability and bifurcation analyses for coupled spatial systems.
\end{abstract}

\keywords{FitzHugh-Nagumo; Mathematical biology; Neuronal dynamics; Cardiac systems; Nonlinear dynamics; Bifurcation analysis; Synchronization; Traveling waves; Spatiotemporal patterns}
\maketitle

\section{Introduction}

Mathematical modeling is a powerful tool used in various fields to study and
understand complex phenomena~\cite{MathematicalModeling}. It involves the use
of mathematical equations and techniques to represent real-world problems,
systems, or processes in a simplified and abstract form. This allows analyzing
and predicting the behavior of the system under different conditions, and
making informed decisions based on the results of the analysis. Mathematical
modeling is widely used in physics~\cite{MathematicalModelingPhysics},
engineering~\cite{MathematicalModelingEngineering},
finance~\cite{MathematicalModelingFinancing},
biology~\cite{MathematicalModelingBiology},
medicine~\cite{MathematicalModelingMedicine1}, and many other
fields~\cite{MathematicalModelingOthers1,MathematicalModelingOthers2}. It has
led to significant advances in scientific research, technological innovation,
and practical problem-solving. In this way, mathematical modeling has become an
indispensable tool in modern science and society.\\

\begin{figure}
\includegraphics[width=\textwidth]{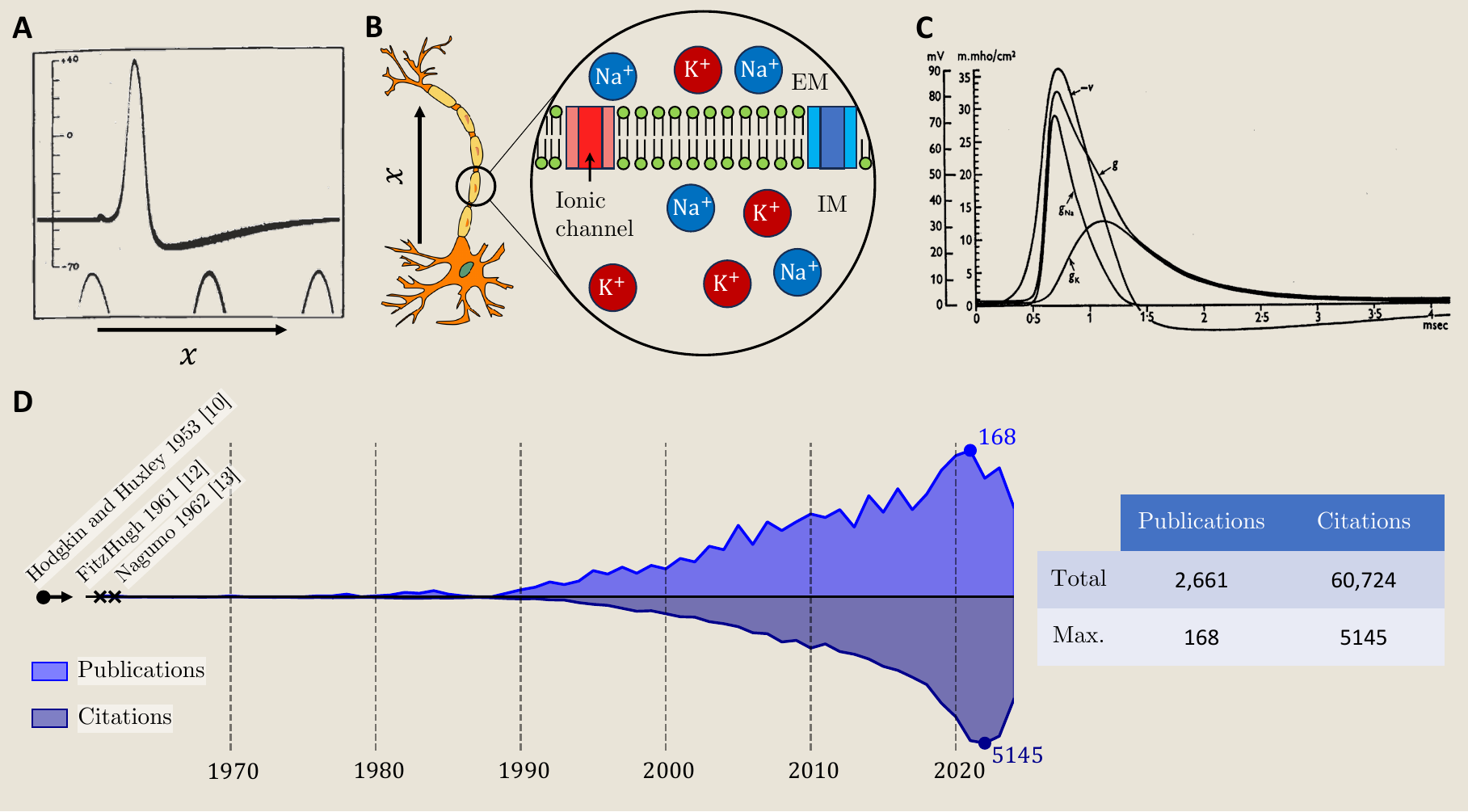}
\caption{\textbf{Exploring neuronal dynamics: The Hodgkin-Huxley model and its
 impact on subsequent research via the FitzHugh-Nagumo model.}
 \textbf{A}. Depiction of the action potential across an axonal
 membrane, adapted from Hodgkin and Huxley's seminal
 work~\cite{hodgkin1939action}, with the potential difference measured
 in millivolts and the external environment set as the zero potential
 reference.
    \textbf{B}. Illustration of ion channels within a neuronal axon, detailing the exchange of sodium and potassium ions between the extracellular medium (EM) and the intracellular medium (IM).
    \textbf{C}. Demonstration of excitable behavior as characterized by the HH
 model, adapted from Fig. 17 from the original
 study~\cite{HodgkinHuxley}. \textbf{D}. Publication and citation data
 gathered from the Web of Science using ``FitzHugh-Nagumo'' as the search
 term (September 19, 2024). Note that these metrics do not include
 papers where the FHN model is not the primary focus, or papers
 published before the 1970s when the model name was first used. This
 shows the impact of the subsequent research using the FHN model as a
 simplified description of the HH model dynamics.
 \label{fig:HHmodel}}
\end{figure}

In this work, we review the FitzHugh-Nagumo (FHN) equation as one successful example of mathematical modeling (Fig.~\ref{fig:HHmodel}D). We will briefly discuss how the FHN model was introduced in the field of neuroscience, becoming a widely used model for neuronal dynamics. However, the FHN model exists in many variations of the original system, and it has been studied in many different fields due to its simplicity and its rich and generic dynamical behavior, which we highlight here. The original FHN equation shows bistability, (relaxation) oscillations, and excitability. When coupling multiple FHN models, for example through spatial diffusive coupling, traveling waves and extended patterns can be found. In this review, we will highlight many of the most commonly found dynamical states and bifurcations.

\subsection{Action potentials in neurons: the Hodgkin-Huxley model}

Information in nerve fibers is encoded through electrical membrane changes, known as action potentials. In 1939, Alan Hodgkin and Andrew Huxley managed to make the first intracellular recording of an action potential by inserting microelectrodes into the giant axons of squid (Fig.~\ref{fig:HHmodel}A). In their experiments, Hodgkin and Huxley were able to demonstrate that this action potential was the result of two distinct contributions: a rapid inward current carried by sodium (Na$^+$) ions, and a more slowly activating outward current carried by potassium (K$^+$) ions.
At rest, a cell is not actively transmitting a signal, and there is a potential difference of about 70~mV across the membrane. When a sufficiently strong external stimulus disrupts the cell, the membrane potential rapidly increases, leading to a phenomenon known as depolarization. Subsequently, the membrane potential decreases back to its baseline level, a phase referred to as repolarization.\\

Using voltage-clamp methods, they discovered that the permeability of the
membrane for sodium and potassium was regulated independently, with the
conductance depending on both time and the membrane potential. Remarkably, they
provided a complete computational model for the action potential in a single
cell, now known as the Hodgkin-Huxley (HH) model~\cite{HodgkinHuxley}, which
was a major advancement of our understanding. For their seminal work, they were
awarded the Nobel Prize in Physiology or Medicine in
1963~\cite{Nobelprize.org1963}. \\

Using the HH equations, Hodgkin and Huxley modeled the measured smooth changes in the current by introducing specific (probabilistic) terms to capture that ion channels can be either open or closed. Indeed, voltage-gated ion channels play a pivotal role in governing the dynamics of action potentials. These channels respond to changes in the nearby electrical membrane potential by undergoing conformational changes, thus controlling their opening and closing. 
Since the lipid bilayer is generally impermeable to ions, the trafficking of ions across the cell membrane is regulated through these voltage-gated ion channels. The HH model elucidates the electrical behavior of ion channels within the cell membrane, specifically addressing the passage of sodium and potassium ions (Fig.~\ref{fig:HHmodel}B). The total ionic current was then represented as the sum of sodium, potassium, and leak currents.\\

The HH model consists of four ordinary differential equations, each corresponding to a state variable: $V_m(t)$, $n(t)$, $m(t)$, and $h(t)$,

\begin{equation}\begin{array}{ @{} l  @{} }        I=C_m\frac{dV_m}{dt}+\overline{g}_Kn^4(V_m-V_k)+\overline{g}_{Na}m^3h(V_m-V_{Na})+\overline{g}_l(V_m-V_l), \vspace{3pt}\\
        \displaystyle\frac{dn}{dt}=\alpha_n(V_m)(1-n)-\beta_n(V_m)n, \vspace{3pt}\\
        \displaystyle\frac{dm}{dt}=\alpha_m(V_m)(1-m)-\beta_m(V_m)m, \vspace{3pt}\\
        \displaystyle\frac{dh}{dt}=\alpha_h(V_m)(1-h)-\beta_h(V_m)h.
\end{array}
\end{equation}

Here, $I$ represents the current per unit area, while $\alpha_{i}$ and $\beta_{i}$ denote voltage-dependent rate constants specific to the $i$th ion channel. $\overline{g}_{n}$ represents the maximum conductance value, and $n$, $m$, and $h$ are dimensionless probabilities ranging between $0$ and $1$. These probabilities correspond to the activation states of potassium channel subunits, the activation of sodium channel subunits, and the inactivation of sodium channel subunits, respectively.\\

This system is four-dimensional and nonlinear, making it analytically unsolvable, with no closed-form solution available. However, numerical simulations enable the exploration of certain properties and general behaviors, such as the existence of oscillations and excitability. In the context of dynamical systems, excitability refers to the property of a system to respond to a stimulus or perturbation by generating a transient activity. Systems exhibiting excitability often display a threshold behavior, where a certain level of input or perturbation is required to trigger a response. In an excitable system, there are typically three phases of activity:\\

\textbf{Resting state}: The system is at rest or in a stable equilibrium.

\textbf{Excited state}: When a stimulus or perturbation surpasses a certain threshold, the system transitions to an excited state. During this phase, the system exhibits a rapid and transient response.

\textbf{Refractory period}: Following the excitation, there is a refractory period during which the system is less responsive and needs time to recover before it can be excited again.\\

The fact that the HH model correctly captures these excitable
dynamics~\cite{HodgkinHuxley} was a major realization as it provided a detailed
understanding of the measured action potential and the dynamical process of
neural excitability (Fig.~\ref{fig:HHmodel}C). 

\subsection{A simplified description of neuronal excitability: the {FitzHugh}-Nagumo model}

While the HH model effectively replicates many neuronal physiological
phenomena, its aforementioned complexity is a significant drawback. In the
1960s, Richard FitzHugh introduced a simplified model to capture the dynamics
of neuronal excitability, which Jinichi Nagumo further refined a year later
(Fig.~\ref{fig:HHmodel}D). This model is now recognized as the FitzHugh-Nagumo
(FHN) model~\cite{1,2}. To develop this model, FitzHugh focused on preserving
important dynamic characteristics of the HH model, namely the presence of
excitability (Fig.~\ref{fig:HHmodel}C) and oscillations. \\

FitzHugh started from the van der Pol oscillator equation~\cite{vanDerPol2},
introduced in the 1920s by Van der Pol~\cite{vanDerPol1}. As its name suggests,
the Van der Pol oscillator admits oscillations which are
\textit{relaxation-like}, meaning they feature periods of slow progress close
to a ``low state'' and a ``high state'' punctuated by rapid transitions between
them. Van der Pol's equation is built from the simple differential equation for
the damped harmonic oscillator by replacing the damping constant with a damping
coefficient that depends quadratically on $x$, thus introducing a
nonlinearity:
\begin{equation}
    x_{tt}+kx_t+x=0\longrightarrow x_{tt}+c(x^2-1)x_t+x=0.
    \label{VanderPol}
\end{equation}

Note that due to this nonlinear damping, there is only effective damping for
small $x$, while for $x^2 > 1$ the nonlinear term describes
amplification rather than damping. To more easily interpret the dynamics of the
van der Pol equation, one can use the Li\'enard's
transformation~\cite{Lienards1}
\begin{equation}
    y = x_t/c + \left(\dfrac{x^3}{3} - x\right),
\end{equation}
leading to a system of two differential equations:
\begin{equation}\begin{array}{ @{} l  @{} }    x_t = c\left[y-\left(\displaystyle\frac{x^3}{3}-x\right)\right], \vspace{3pt}\\
    y_t = -\displaystyle\frac{1}{c} x,
\end{array}
\end{equation}
from which one can clearly see the separation in time scales of both equations. While the first one evolves fast on $O(c)$, the second one is much slower on the order of $O(c^{-1})$. Building on the van der Pol oscillator, FitzHugh modified the equations as follows:
\begin{equation}\begin{array}{ @{} l  @{} }        x_t=c \left[y-\left(\displaystyle\frac{x^3}{3}-x\right)+z\right], \vspace{3pt}\\
        y_t=-\displaystyle\frac{1}{c}(x-a+by).
\end{array}
\label{EqFitzHugh}
\end{equation}

\begin{figure}
\includegraphics[width=\textwidth]{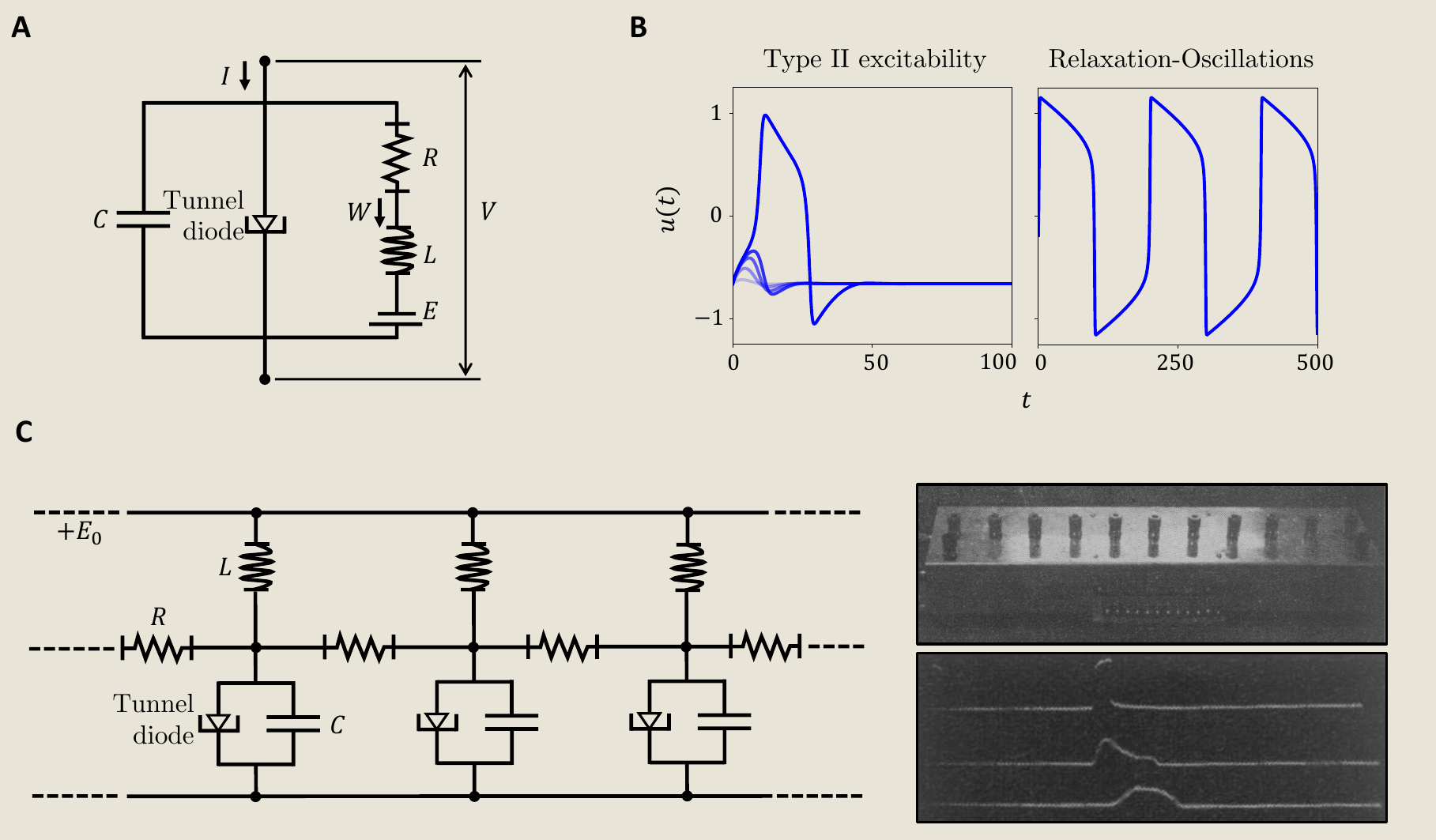}
\caption{\textbf{Electrical analog of the FitzHugh-Nagumo model}
    \textbf{A}. Circuit representation: The FHN model's electrical counterpart includes key components that mimic biological neuronal dynamics. This circuit comprises:
(i) A capacitor, symbolizing the neuronal membrane's capacitance $C$.
(ii) A tunnel diode, representing the nonlinear ionic current $F(V)$.
(iii) A resistor, indicative of the channel resistance $R$.
(iv) An inductor $L$ and a battery $E$, completing the circuit to model the rest of the system's dynamics.
\textbf{B}. Neuronal dynamics regimes: Through simulations of Eq.~\ref{EqTemporal}, we capture essential neuronal behaviors:
(i) Excitable regime: At $(a,b,\varepsilon)=(0.1,1.5,0.1)$, the circuit mimics the excitable nature of neurons, responding robustly to stimuli beyond a certain threshold.
(ii) Oscillatory regime: With parameters $(a,b,\varepsilon)=(0,0.5,0.01)$, the system exhibits periodic oscillations, typical of active neuronal firing patterns.
\textbf{C}. Spatial coupling in neuronal arrays: Extending the model to
 encompass spatial interactions involves linking multiple such circuits.
 This approach, as demonstrated in J. Nagumo \textit{et al.}'s seminal
 work~\cite{2}, allows for the exploration of wave propagation and
 collective behaviors in a network of neuron-like elements, mirroring
 the complex dynamics observed in biological neural networks.\label{fig:FHNmodel}}
\end{figure}

FitzHugh introduced the parameter $z$ to represent membrane current density, the variable $x$ is related to the membrane voltage and the sodium activation, and the variable $y$ corresponds to the sodium inactivation and the potassium activation. In this context, these equations thus work as an activator-inhibitor model. Later, Nagumo, Arimoto, and Yoshizawa proved the equivalence of this model with an electrical circuit.
 Fig.~\ref{fig:FHNmodel}A shows this electrical equivalent, which includes a capacitor (representing membrane capacitance $C$), a tunnel diode [$F(V)$], a resistor (representing channel resistance $R$), an inductor ($L$), and a battery ($E$). The dynamics of this electrical circuit are captured by the following equations:
\begin{equation}\begin{array}{ @{} l  @{} }        C V_t = I - F(V) - W, \vspace{3pt}\\
        L W_t = E - R W + V,
\end{array}
\end{equation}
where $V$ is the voltage across the circuit, $W$ corresponds to the current through the $R-L-E$ branch of the circuit, and $F(V)$ is the current flowing through the tunnel diode driven by the voltage $V$. When the nonlinear function $F(V)$ is cubic, this set of equations can be readily transformed into Eqs.~\ref{EqFitzHugh}.\\

The model Eqs.~\ref{EqFitzHugh} are invariant under the transformation $x\to-x$, $y\to-y$ and $a\to-a$, so considering $a>0$ is sufficient to understand its dynamics. The equations are also invariant under the transformation $(t,c)\to(-t,-c)$, so we can also restrict ourselves to $c>0$. To cast these equations in a simpler form, we consider the transformation $t \rightarrow c t$ and $\varepsilon=1/c^2$. We also absorb the parameter $z$ into the variable $y$ ($y \rightarrow y + z$), such that we redefine $a \rightarrow a-bz$:
\begin{equation}\begin{array}{ @{} l  @{} }        x_t=-\displaystyle\frac{x^3}{3}+x+y,\vspace{3pt}\\
        y_t=\varepsilon(x+by+a).
\end{array}
    \label{EqTemporal_Og}
\end{equation}

Here in this review, we will consider $x\equiv u$ and $y\equiv -v$, which are most commonly used in the literature to represent the FHN model. Additionally, without the loss of generality in its dynamics, we omit the factor 3 in the first equation:
\begin{equation}\begin{array}{ @{} l  @{} }        u_t=-u^3+u-v, \vspace{3pt}\\
        v_t=\varepsilon(u-bv+a),
\end{array}
    \label{EqTemporal}
\end{equation}

In this most basic form, the FHN model thus consists of two coupled, nonlinear ordinary differential equations, where the first one describes the fast evolution of the membrane voltage of a neuron ($u$), while the second one represents the slow recovery through the opening of potassium channels and the inactivation of sodium channels ($v$).

\subsection{Excitability in the generalized {FitzHugh}-Nagumo model}

As observed in  Fig.~\ref{fig:FHNmodel}B, the main dynamics of the HH model are also observed in the FHN model, i.e.~excitability and  \textit{relaxation-like} oscillations. Excitability is classified as \textit{type II}, characterized by the lack of a distinct threshold. Excitability and oscillatory dynamics are essential to model neuronal spiking, and although the quantitative behavior may change, the qualitative behavior is preserved by the simplified model. The key ingredients that preserve these dynamical behaviors are the time scale separation and the cubic shape of the first differential equation. This separation implies the presence of fast and slow regions (quick versus slow changes in  Fig.~\ref{fig:FHNmodel}B), while the influence of the latter will be explained in more detail in Fig.~\ref{timeScaleSeparation}. \\

In the literature, different variants of the FHN model have been used, that all fall under the following \textit{generalized} FHN equations:
\begin{equation}\begin{array}{ @{} l  @{} }        u_t=f(u,v),\vspace{3pt}\\
        v_t=\varepsilon g(u,v),
\end{array}
\end{equation}
where $f(u,v)$ is a nonlinear function of $u$ and a linear function
of $v$, $g(u,v)$ a linear function of both $u$ and
$v$, and $\varepsilon$ represents the time scale separation between both
equations. The nonlinearity of $f(u,v)$ is typically represented by a cubic
function, written as $-u^3+u$ as in Eq.~\ref{EqTemporal} or factorized as
$u(1-u)(u-c)$, which is equivalent because the quadratic term can be eliminated
as shown in Rocsoreanu's book~\cite{5}. The cubic shape of $f(u,v)$ is also
often approximated piecewise-linearly using $f(u,v) = -H(u-a)+u-v$ with $H$ a
Heaviside function, which was first proposed by McKean in the 1970s~\cite{16}.

\subsection{Spatial propagation of action potentials in coupled {FitzHugh}-Nagumo models}

While the excitable and oscillatory behavior in the FHN equation capture typical neuronal dynamics (Fig.~\ref{fig:FHNmodel}B), in reality, the action potentials also propagate along the axon. Therefore, diffusion in space was added to both the HH model and the FHN model:
\begin{equation}\begin{array}{ @{} l  @{} }        u_t=D_u\Delta u-u^3+u-v, \vspace{3pt}\\
        v_t=D_v\Delta v+\varepsilon(u-bv+a).
\end{array}
    \label{EqSpatial}
\end{equation}

In the presence of diffusive coupling, traveling wave solutions exist capturing
the spatial propagation of action potentials. In the diffusively-coupled
equations above, we have considered the most general case where the diffusion
coefficients of $u$ and $v$ are different. However, space can
be renormalized such that one coefficient is absorbed into the Laplacian.
Having two diffusion coefficients of similar strength (both $O(1)$) is
typical in chemical reactions. Nevertheless, in neural dynamics and
electrophysiological applications in general it is common to only have
diffusion in the $u$ variable (with $D_v=0$), as done in the
original formulation introduced by Nagumo \textit{et al.}~\cite{2}. \\

\begin{figure}
\includegraphics[width=\textwidth]{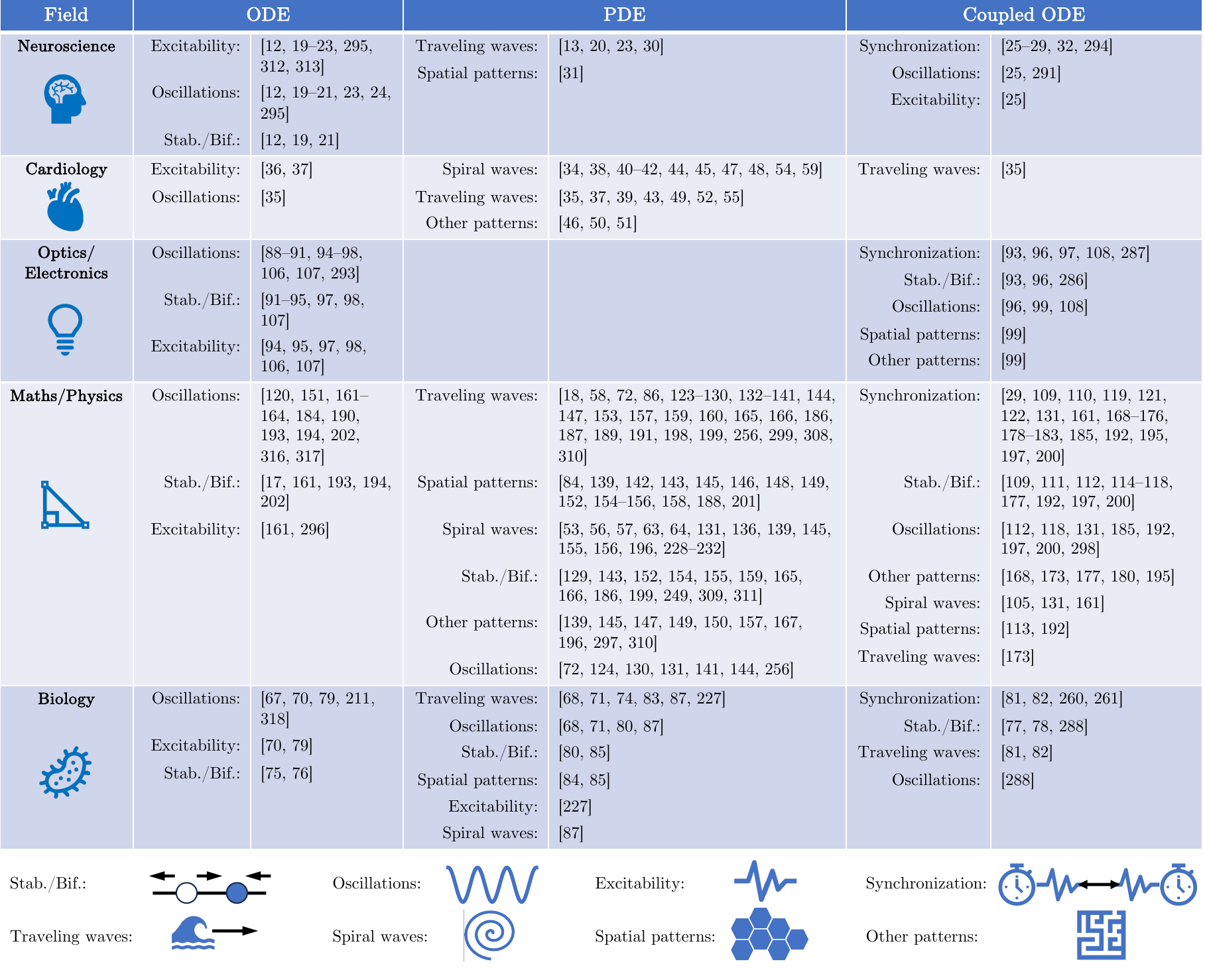}
\caption{\textbf{Overview of FitzHugh-Nagumo model applications}
This table organizes selected literature related to different formulations of the FitzHugh-Nagumo equations, as outlined in Eq.~\ref{EqTemporal}, Eq.~\ref{EqSpatial}, and Eq.~\ref{EqCoupled}. Each listed study is categorized by its application domain and primary focus. It is important to note that, alongside the conventional FitzHugh-Nagumo models, certain studies incorporate variations like piecewise linear models or three-dimensional adaptations to meet particular investigative requirements.\label{fig:PreviosReferences}}
\end{figure}

Nagumo \textit{et al.}  also built an electrical circuit consisting of many sequentially coupled units as illustrated in  Fig.~\ref{fig:FHNmodel}C (left), leading to a system capable of simulating the propagation of action potentials along a nerve axon (Fig.~\ref{fig:FHNmodel}C, right). Of course, this represents an approximation of the diffusively coupled system, as there are only a discrete number of components that are spatially coupled. In fact, one can couple such components governed by the FHN system in a variety of different ways, leading to the following system of discretely coupled FHN equations, that is also often studied in the literature:
\begin{equation}\begin{array}{ @{} l  @{} }        {u_i}_t=-u_i^3+u_i-v_i+\sum_j^np(u_j,v_j), \vspace{3pt}\\
        {v_i}_t=\varepsilon(u_i-bv_i+a)+\sum_j^nq(u_j,v_j).
\end{array}
\label{EqCoupled}
\end{equation}
This discretely coupled model configuration is extensively studied in neural dynamics applications (Fig.~\ref{fig:PreviosReferences}). Its formulation allows to observe the effect of noise on a set of neurons and the synchronization between them. Despite its usefulness, the equations are much more complex and the difficulty of the stability analysis is significantly increased. Moreover, the models' behavior heavily depends on the coupling's functional form, which can vary in directionality (bidirectional or unidirectional) and timing (instantaneous or with time delay). 

\subsection{Beyond neuroscience: diverse applications of the {FitzHugh}-Nagumo model}

Despite its simplicity, the FHN model captures many of the essential features
of neuronal dynamics, and has been used to study a wide range of phenomena in
neuroscience~\cite{1,3,4,15,29,77,95,7,96,128,140,182,2,62,63}, including the
synchronization of neural activity~\cite{7,96,128,140,182,al2024criticality},
the dynamics of neural networks~\cite{1,3,15}, and the emergence of complex
spatio-temporal patterns in neural systems~\cite{63} (Fig.~\ref{fig:HHmodel}D).
However, thanks to its generic dynamical behavior, the FHN model finds
applications in various fields beyond neuroscience. In
Fig.~\ref{fig:PreviosReferences}, we provide an overview of the main fields where
the FHN model has helped to advance our understanding of those systems. We
highlight electronics, optics, and biological systems. Given its significant
role in cardiac modeling, the FHN model's application to cardiac systems is
discussed separately. Motivated by its broad applicability and generic
dynamics, the FHN system has triggered a large body of theoretical work,
characterizing bifurcations and its various dynamical behaviors.
Fig.~\ref{fig:PreviosReferences} illustrates the work in the following domains:\\

\textbf{Neuroscience}: The FHN model, initially developed in neuroscience, has
played a pivotal role in understanding the complex dynamics of neurons and
neural
networks~\cite{1,3,4,15,29,77,95,7,96,128,140,182,2,62,63,al2024criticality}.
By describing complex neuronal behaviors with simple equations, the FHN model
has made it easier to analyze and simulate neural dynamics. It offers a
comprehensive view of solutions, providing geometric insights into crucial
biological phenomena like neuronal excitability and spike generation
mechanisms. \\

Through bifurcation analysis, researchers have gained insights into the emergence and transitions between various realistic dynamics, spanning excitability, oscillations, and bistability. Focusing on key aspects such as excitation and inhibition, the FHN model offers a practical approach to explore the fundamental properties of neurons. Additionally, its extension to spatially distributed systems enables the modeling of action potential propagation along axons. \\

Moreover, the FHN model has proven invaluable in studying neuronal interactions within networks. By linking multiple FHN units, researchers have gained deeper insights into network communication, synchronization, and the generation of coordinated activity patterns. This has greatly enhanced our understanding of phenomena like neuronal oscillations and information processing in the brain. 
Furthermore, the FHN model has laid the groundwork for the development of more sophisticated neuronal models, serving as a platform for incorporating additional biological complexity. By leveraging insights from the FitzHugh-Nagumo framework, researchers continue to refine our understanding of neuronal function and dysfunction in both health and disease.\\

\textbf{Cardiology}: In 1962, based on available experimental data, Noble
adapted Hodgkin and Huxley's model of excitable neurons to describe the
dynamics of Purkinje fibers in the heart~\cite{HH_cardiac1}. The key
modification to the HH model involved considering potassium current flowing
through two nonlinear resistances. Since then, a plethora of cardiac models
have been developed, ranging from subcellular to whole organ descriptions, with
varying levels of detail~\cite{Alonso_2016}.\\

Among the various models used to study cardiac cells, the FHN model stands out
as one of the simplest yet most widely studied for capturing the general
dynamical features of cardiac
cells~\cite{71,78,79,81,14,66,67,79,80,84,93,102,105,124,130,159,166,177,180,Alonso_2016,bar1993turbulence,doi:10.1126/science.1080207,ben2024stability}.
In addition to the FHN model, several other generic models are used to
reproduce the qualitative properties of cardiac excitation.
The Barkley model, which is very similar to the FHN model, is widely used in
numerical studies due to its efficient numerical
implementation~\cite{doi:10.1142/S0218127497001692,BARKLEY199161}. The
B\"ar-Eiswirth model extends the Barkley model and was introduced specifically
to study spiral breakup~\cite{bar1993turbulence}. Similarly, the Puschino model
is extensively utilized in numerical simulations of excitable
media~\cite{PANFILOV1987215}.
The Karma model, another variant, was developed to demonstrate spiral wave
breakup due to alternans~\cite{PhysRevLett.71.1103}. It generates an action
potential with a more realistic fast upstroke and slow recovery compared to the
standard FHN model. Additionally, the Aliev-Panfilov model adapts the FHN
framework to capture the rate-dependent variations in action potential
duration~\cite{80}.\\

In cardiac systems, many models are used to investigate the existence and
dynamics of spirals, which are wave phenomena arising from anisotropy or
spatial defects and characterized by a core and rotation
frequency~\cite{mathematicalSpirals}. Electrical impulses, denoted by the
variable $u$, propagate in a wave-like manner in the heart, and it has
been observed that ventricular fibrillations result from the collision of two
spirals rotating in opposite directions~\cite{cardiacSpirals1,cardiacSpirals2}.
The FHN model is frequently utilized to simulate such behavior and analyze the
effects of spiral breakup~\cite{66,69,78,79,68,14,80,81}.\\

Furthermore, the heart is often analyzed as a system of discretely coupled
oscillators, reflecting its compartmental structure. Certain oscillatory
compartments have been described using FHN models or variants
thereof~\cite{cardiacCoupled1,cardiacCoupled2}. For instance, Grasman
considered van der Pol oscillators, which can be replaced by FHN oscillators,
to construct a system of three coupled oscillators representing the sinoatrial
node, atrium, and ventricle, with a delay between the latter
two~\cite{cardiacCoupled2}.\\

\textbf{Biology}: While the FHN model is predominantly utilized in describing
neuronal and cardiac systems, its application extends to various other
biological contexts. The FHN equations offer a simplified yet versatile
representation of interlinked positive and negative feedback loops, capable of
generating diverse dynamical responses including switches, pulses, and robust
oscillations~\cite{tsai2008robust}. Furthermore, when coupled with diffusion,
the FHN model can produce trigger waves that rapidly propagate these dynamic
behaviors over large distances~\cite{63,64,ditalia2022}.\\

In the context of cellular processes, the FHN model has been employed to
describe the oscillatory dynamics of the activity of the cyclin
B-cyclin-dependent kinase 1 (Cdk1) complex during the cell division
cycle~\cite{13,64,74,75,189,tsai2008robust}. Additionally, it has been
instrumental in studying oscillations in the concentration of free cytosolic
Ca$^{2+}$, a crucial cellular control mechanism~\cite{173}.
Beyond cellular biology, the FHN model has found applications in diverse
organisms. In bacteria, it has been used to investigate polarization
dynamics in response to electrical stimuli, akin to neuronal
responses~\cite{156}. In plants, FHN-like models have served as cell signaling
models for pulse-like jasmonate responses~\cite{143}.\\

The coupling of FHN units in networks has shed light on various physiological
phenomena. For instance, in mammalian pancreas $\beta$ cells, coupled FHN
units elucidate how global oscillations emerge despite the absence of a
pacemaker region~\cite{178,scialla2021hubs}. Similarly, in plant systems,
coupling FHN models with photosynthesis models aids in understanding the impact
of environmental variables such as temperature~\cite{144}.
Spatially extended systems described by diffusively coupled FHN equations have
provided insights into wave propagation during mitosis in \textit{Xenopus} frog
embryos~\cite{64,74,75,189}, as well as calcium waves triggered by
fertilization~\cite{174}. The FHN model has also been applied to study
propagating action potentials in vascular plant tissues~\cite{145,146} and
morphoelastic waves driving the locomotion of soft robots~\cite{147}.
In ecology, adding cross-diffusive effects to the FHN model has enabled studies
on pattern formation in living systems, such as chemotactic movement in
\textit{Escherichia coli} and interactions in predator-prey
dynamics~\cite{148,149,151,175}.\\

\textbf{Electronic and optical systems}: The significance of understanding the
dynamics of neurons and neural networks has long been acknowledged, both for
unraveling the complexities of the human brain and for applying analogous
concepts to machine learning applications. Consequently, researchers have
endeavored to emulate neuron cells using electronic circuits. Many of these
circuits have been specifically devised to replicate the dynamics of the FHN
model~\cite{132,133,134,136,137,138,139,158,161,169,170,176}.\\

In recent years, advancements in integrating optoelectronic components onto
photonic integration platforms have spurred extensive investigations into
ultrafast artificial neural networks for information
processing~\cite{shastri_photonics_2021}. All-optical spiking neurons have been
successfully demonstrated using semiconductor
lasers~\cite{PhysRevA.82.063841,PhysRevE.84.036209,PhysRevLett.112.183902,6497478}.
Furthermore, neuronal dynamics described by the FHN system have been realized
through (electro-)optical implementations~\cite{172,110,113,huang4946665phase}.\\

\textbf{Mathematics and physics}: While the FHN model was specifically
developed to describe neuronal dynamics, it has served more generally as a
cornerstone in understanding the rich dynamical behaviors exhibited by
excitable systems through dynamical systems
analysis~\cite{5,8,9,10,11,12,19,20,21,22,23,24,25,26,27,30,31,32,33,34,35,36,37,38,39,40,41,42,44,45,46,47,48,49,50,51,53,54,55,56,57,59,60,65,72,73,75,83,85,86,87,89,90,91,92,94,97,98,99,101,103,104,107,108,111,114,115,116,117,118,120,121,122,123,125,126,131,135,141,148,150,151,152,153,154,155,157,162,163,164,165,171,172,179,181,182,183,185,186,187}.
By representing the essential characteristics of excitable cells, the FHN model
provides a simplified yet powerful framework for exploring a wide range of
dynamical phenomena. Through bifurcation analysis, researchers have uncovered
complex patterns of behavior emerging from the model's equations. These
analyses reveal how changes in parameters, such as excitability thresholds or
coupling strengths, can lead to transitions between different dynamical
regimes, including fixed points, limit cycles, and chaotic behavior. Moreover,
stability analysis techniques enable the identification of critical points
where small perturbations can lead to qualitatively different system behaviors,
shedding light on the robustness and sensitivity of excitable systems.\\

Furthermore, phase-plane analysis has been instrumental in elucidating the dynamics of the FHN model by visualizing trajectories in its phase space. By plotting the evolution of variables such as membrane potential and recovery variable, one can gain insights into the underlying mechanisms driving excitability, oscillations, and wave propagation. Through phase-plane analysis, the FHN model's behaviors, such as action potential generation and refractoriness, can be understood in terms of the system's underlying dynamics. Additionally, techniques such as nullclines analysis provide a geometric understanding of the model's behavior, revealing regions of parameter space associated with different dynamical regimes. \\

 \textbf{Other}: Additional examples not mentioned previously encompass
chemical reaction dynamics~\cite{167,171,183}, the dynamics of elastic
excitable media~\cite{160}, or the study of computational
algorithms~\cite{6,52,61,82}. In exploring chemical reaction dynamics, the FHN
model has been instrumental in investigating the emergence of chemical
turbulence through reaction and diffusion processes~\cite{167,183}. One
mechanism for generating chemical turbulence involves spiral waves that undergo
breakup, a phenomenon initially demonstrated in the context of the
Belousov-Zhabotinsky chemical reaction-diffusion
system~\cite{ouyang_transition_1996}, and subsequently studied extensively
within the framework of the FHN model~\cite{183}. Furthermore, the FHN system
has been employed to understand spatial patterns in a forced FitzHugh-Nagumo
reaction-diffusion model, reproducing dynamics akin to those observed in the
Belousov-Zhabotinsky system~\cite{171}.\\

In the context of elastic excitable media, the FHN-equivalent Burridge-Knopoff
model has been applied to characterize the frictional sliding dynamics in
earthquake fault systems influenced by viscous friction~\cite{160}. Due to the
limited number of found references in these fields, they were not included as a
separate category in the table depicted in  Fig.~\ref{fig:PreviosReferences}.\\

In the field of computational algorithm development and analysis, the FHN model
has also proven valuable. It serves not only as a framework to investigate
specific dynamics like traveling waves~\cite{6} and Turing pattern
formations~\cite{52} but also as a versatile tool for examining how algorithms
perform across different boundary shapes and conditions~\cite{61,82}.

\subsection{Outline}

In this review, we aim to offer an in-depth exploration of the diverse dynamical behaviors encapsulated within the FHN model. The widespread adoption of the FHN model across physics and biology can be attributed to the model's remarkable versatility in capturing a wide array of dynamical phenomena while maintaining a relatively simple mathematical formulation.\\

Our review is structured around delineating the most prominent dynamical behaviors observed within the FHN model. We categorize our analysis into three primary sections: (i) examining the foundational FHN model, characterized by a system of two nonlinear coupled ordinary differential equations (ODEs) [Eq.~\ref{EqTemporal}]; (ii) studying the diffusively coupled FHN model, which introduces spatial coupling through diffusion [Eq.~\ref{EqSpatial}]; and (iii) exploring discretely coupled FHN equations [Eq.~\ref{EqCoupled}]. In each section, we complement our discussion of observed dynamics with thorough stability analyses and bifurcation studies. This approach allows readers to navigate the parameter space effectively, enabling them to target specific dynamical regimes of interest.\\

We anticipate that our review will serve as a valuable resource for modelers and experimentalists across various disciplines outlined in  Fig.~\ref{fig:PreviosReferences}, where the FHN model has demonstrated utility. These fields span neuroscience, cardiac dynamics, other biological processes, electronics and optics, and chemistry. Furthermore, we hope that our exploration will inspire the application of the FHN model in additional systems exhibiting analogous dynamics, thereby expanding its scope and impact across diverse scientific domains.

\section{Dynamics in the {FitzHugh}-Nagumo model}
\label{temporal}

\begin{figure}
\includegraphics[width=\textwidth]{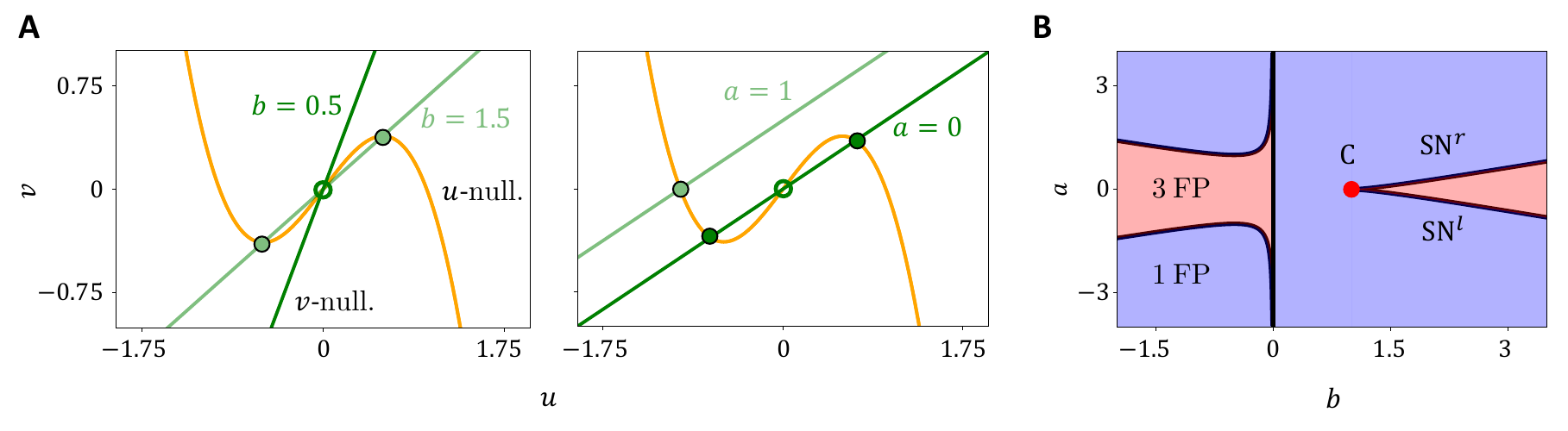}
\caption{\textbf{Fixed points in the FHN model.} \textbf{A}. The nullclines in the ($u,v$) phase space ($\varepsilon=0.01$) and the effect of the parameters $a$ and $b$ on them are shown. \textbf{B}. The saddle node bifurcation in parameter space ($\varepsilon=0.01$) is shown and the mono- (1 FP) and tri-valued (3 FP) regions are differentiated. SN indicates the saddle node bifurcations and their intersection in the cusp (C) bifurcation.\label{fig:fixedPoints}}
\end{figure}

The basic FHN model [Eq.~\ref{EqTemporal}] has been studied extensively
because, despite its apparent simplicity, it exhibits complex dynamics and
bifurcations. Here, we will limit ourselves to providing an overview of the
most widely used dynamical regimes, such as monostability, multistability,
relaxation oscillations, and excitability. For an in-depth bifurcation
analysis, the reader is referred to the comprehensive work by Rocsoreanu
\textit{et al.}~\cite{5}.

\subsection{Stationary solutions}

The simplest solutions of Eq.~\ref{EqTemporal} are the stationary states, also known as fixed points (FP) or equilibrium points.
These points satisfy the condition where the time derivatives of both variables are zero [i.e.,~$(u_{\text{FP}_t},v_{\text{FP}_t})=(0,0)$], leading to a set of equations that define the so-called \textit{nullclines} in the $(u,v)$ phase plane (Fig.~\ref{fig:fixedPoints}A):
\begin{equation}\begin{array}{ @{} l  @{} }        0=-u^3+u-v,\vspace{3pt}\\
        0=\varepsilon(u-bv+a).
\end{array}
    \label{EqNC}
\end{equation}

At these nullclines, one of the variables remains constant over time: either $u$ or $v$ does not change in time. The intersections of the nullclines represent the stationary solutions $(u,v)=(u_{\text{FP}},v_{\text{FP}})$, which are solutions to a cubic equation in terms of one of the variables ($u_{\text{FP}}$):
\begin{equation}
bu_{\text{FP}}^3+(1-b)u_{\text{FP}}+a=0,
\label{EqBaseI}
\end{equation}
with $v_{\text{FP}}=(u_{\text{FP}}+a)/b$. This cubic equation typically does not have a straightforward analytical solution, but its roots can indicate the existence of either one or three coexisting stationary states, depending on the system parameters. In the latter case, we label those solutions $u_{\text{FP}}^b$, $u_{\text{FP}}^m$, and $u_{\text{FP}}^t$. As a result, the parameter space is divided into regions of monostability and multistability, delineated by fold lines that correspond to the conditions where the system transitions between having a single fixed point and multiple fixed points. These transitions are associated with saddle-node (SN) bifurcations, also called fold bifurcations, and the specific conditions under which they occur depend on the system parameters. These folds are located at $(u,a)=(u_{\text{SN}}^{\pm},a_{\text{SN}}^{\pm})$ where
\begin{equation}
\label{folds}
    u_{\text{SN}}^{\pm}=\pm\sqrt{\frac{b-1}{3b}},  a_{\text{SN}}^\pm=\frac{2}{3}(1-b)\sqrt{\dfrac{b-1}{3b}},
\end{equation}
which are defined for $b\in (-\infty,0)\cup[1,\infty)$. The fixed points are triple-valued when $u_{\text{SN}}^-<u_{\text{FP}}<u_{\text{SN}}^+$, and single-valued otherwise. The presence of these saddle-node bifurcations and their implications for the system's dynamics are illustrated in  Fig.~\ref{fig:fixedPoints}A, which shows how variations in the parameters affect the number and positions of the stationary states.

Figure~\ref{fig:fixedPoints}B further explores the parameter space ($a,b$), highlighting the regions associated with different dynamical behaviors and the critical points where these behaviors change. The fold lines $a=a_{\text{SN}}^\pm(b)$ define the regions of monostability (1 FP) and multistability (3 FP). For $b<0$, the system always shows a region of multistability. For $b>0$, multistability arises above the cusp bifurcation occurring at $b=1$ (C) where the fold points are created.

\subsection{Linear stability of the stationary solutions}
\label{lin_sta_temporal}

To determine the linear stability of the fixed points, we introduce small
perturbations of the form $(u,v)=(u_{\text{FP}},v_{\text{FP}})+\epsilon(\xi_u,\xi_v)e^{\sigma t}+c.c$., with
$\epsilon\ll1$~\cite{guckenheimer_nonlinear_1983}. A fixed point is considered
stable if ${\rm Re}[\sigma]<0$, and unstable if ${\rm Re}[\sigma]>0$. The condition
${\rm Re}[\sigma]=0$ thus marks the occurrence of a local bifurcation. To obtain
$\sigma$, we compute the Jacobian matrix $J$ associated with the
linearization of Eq.~\ref{EqTemporal} around the equilibrium points and
solve the linear eigenvalue problem
\begin{equation}
    (J-\sigma I_{2\times2})
\begin{pmatrix}
        \xi_u\\
        \xi_v
\end{pmatrix}
    =
\begin{pmatrix}
        0\\
        0
\end{pmatrix}
, J\equiv  
\begin{pmatrix}
        1-3u_{\text{FP}}^2  & -1\\
        \varepsilon & -\varepsilon b 
\end{pmatrix}
.
\end{equation}

The solutions for the eigenvalues are
\begin{equation}
\label{sigma}
 \sigma =  \frac{{\rm Tr}(J) \pm \sqrt{{\rm Tr}(J)^2 - 4 {\rm Det}(J)}}{2},
\end{equation}
where ${\rm Tr}(J)$ is the trace and ${\rm Det}(J)$ the determinant of $J$,
defined by
\begin{equation}\begin{array}{ll}    {\rm Det}(J) = \varepsilon(3bu_{\text{FP}}^2-b+1),\vspace{3pt}\\
    {\rm Tr}(J) = 1-3u_{\text{FP}}^2-\varepsilon b.
\end{array}
\label{EqTraceDeterminant}
\end{equation}

Depending on the eigenvalues, two types of bifurcations may occur. If ${\rm Re}[\sigma]=0$ and ${\rm Im}[\sigma]=0$, the system undergoes a saddle-node bifurcation [Eq.~\ref{folds}]. If ${\rm Re}[\sigma]=0$ and ${\rm Im}[\sigma] \neq 0$, a Hopf bifurcation occurs, leading to self-sustained oscillations. These conditions for limit cycle oscillations imply that ${\rm Tr}(J)=0$ and ${\rm Det}(J)>0$ which gives the following Hopf bifurcation point:
\begin{equation}
  u_{\rm H}=\pm\sqrt{\frac{1-\varepsilon b}{3}},    a_{\rm H}=\pm\dfrac{1}{3}(\varepsilon b^2+2b-3)\sqrt{\dfrac{1-\varepsilon b}{3}},
  \label{trCondition}
\end{equation}
 and the frequency of the nascent oscillation is:
\begin{equation}
    \sigma=\pm i\sqrt{{\rm Det}(J)}=\pm i\sqrt{\varepsilon(1-\varepsilon b^2)}. 
    \label{detCondition}
\end{equation}    
 Equations~\eqref{trCondition} and~\eqref{detCondition} provide the conditions for the Hopf bifurcation. The zero trace condition implies $b \leq 1/\varepsilon$, whereas the positive determinant condition requires $b < 1/\sqrt{\varepsilon}$.

The position of these bifurcations vary as a function of the control parameters of the system $a,b$ and $\varepsilon$, leading to the emergence of different dynamical behaviors and regimes. An example of the distribution of these regimes and the bifurcations which define them is illustrated in the $(a,b)$-phase diagram of  Fig.~\ref{fig:linearStability}B for $\varepsilon=0.01$.

\subsection{Dynamical regimes}

The FHN model is well known for exhibiting three main dynamical behaviors: oscillations, excitability, and bistability, as evidenced in the literature (Fig.~\ref{fig:PreviosReferences}).

\begin{figure}
\includegraphics[width=\textwidth]{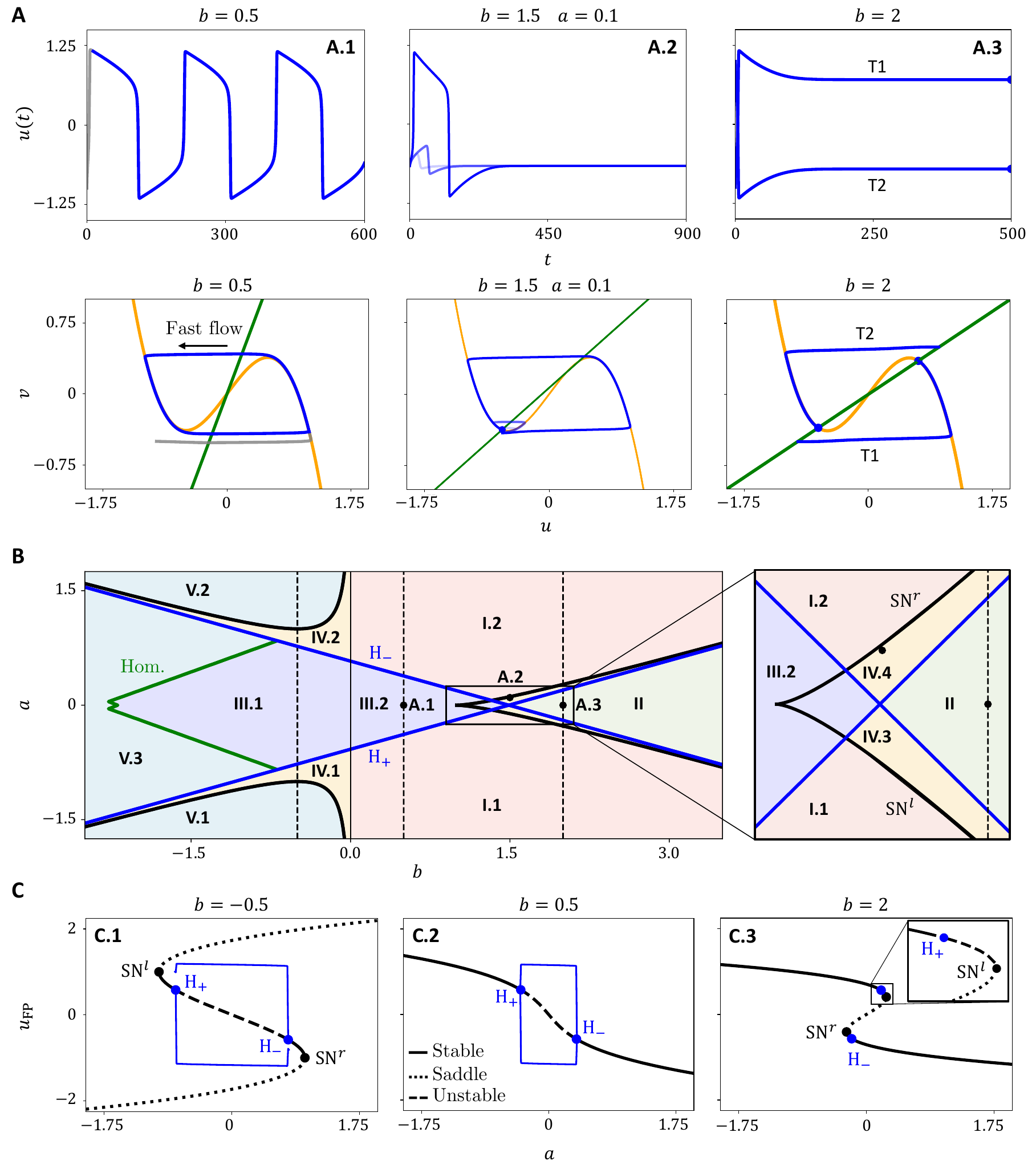}
\caption{\textbf{Dynamics and bifurcations of the FHN model for} $\varepsilon=0.01$. \textbf{A}. Representative time series and phase space dynamics showing relaxation oscillations ($b=0.5$), excitability ($b=1.5$ and $a=0.1$), and bistability ($b=2$). \textbf{B}. Analytically derived Hopf and saddle node bifurcations, along with homoclinic bifurcations determined numerically, delineating various dynamical regions in the ($a,b$) parameter space. An expanded view of a specific area, marked by a black rectangle, is shown on the right. Black dots correspond to parameters used in \textbf{A}, and dashed lines mark sections analyzed in \textbf{C}. \textbf{C} Bifurcation diagrams of three representative regions. The diagram on the left (\textbf{C.1}) hints at an imminent homoclinic bifurcation due to the close proximity of the limit cycle to the saddle-node. The right diagram (\textbf{C.3}), on the other hand, reveals a shift from stability to instability, lacking an associated limit cycle, suggesting a global bifurcation has taken place. The extrema of limit cycle oscillations are shown by the solid gray line.\label{fig:linearStability}}
\end{figure}

\textbf{Relaxation oscillations}.  Characterized by limit cycles, i.e.,~closed-orbit attractors in the phase space, which correspond to oscillations
operating on distinct time scales -- one fast and the other slow.  An example of
this nonlinear oscillation can be observed in  Fig.~\ref{fig:linearStability}A.1
(top) for parameters $a=0$ and $b=0.5$, marked by $\bullet$ in
Fig.~\ref{fig:linearStability}B. In the phase diagram (Fig.~\ref{fig:linearStability}A,
bottom), the oscillation traces a loop around an unstable equilibrium, the
point where nullclines intersect. This oscillatory behavior is crucial for
modeling biological rhythms, including mitotic~\cite{64}, calcium~\cite{17} and
cardiac oscillators~\cite{14}, underscoring the FHN model's applicability in
simulating fast transitions between active and quiescent states in various
biological systems.\\

\textbf{Excitability}. This behavior is intrinsically linked to the oscillatory
phenomena mentioned above and pertains to the system's response to the perturbation
of a fixed point. The magnitude of the perturbation determines the system's
response. In the FHN framework, excitability is for example found for parameters
$a=0.1$ and $b=1.5$, as illustrated in
Fig.~\ref{fig:linearStability}A.2. Minor perturbations result in a swift return to
the baseline state. Conversely, large perturbations drive the system along a
nontrivial trajectory (excursion) in the phase space before resettling to the
resting state. This trajectory, associated with a `spike' in temporal dynamics,
is invariant to the perturbation's specifics. Excitability is the most used
regime in neuroscience and it can reproduce neuronal spikes in the non-coupled
cases~\cite{13,1,3} or in coupled cases~\cite{2}.\\

\textbf{Bistability}. A nonlinear system exhibits bistability when two stable
fixed points coexist for the same set of parameters. This perhaps less frequently
examined behavior within the FHN model is depicted in
Fig.~\ref{fig:linearStability}A.3 for parameters $a=0$ and $b=2$. The
phase space diagram illustrates how different initial conditions converge to
separate stable fixed points, located where the nullclines intersect.
Bistability contributes to the emergence of phenomena such as relaxation
oscillations, excitability, and hysteresis~\cite{bisToOsc1,biToOsc2,bisToOsc3},
vital for the model's relevance and applicability.\\

The complexity of the dynamics within the FHN model requires a more detailed analysis, as depicted in  Fig.~\ref{fig:linearStability}B. By examining cross-sections of this figure at constant $b$ values, we can generate 1D bifurcation diagrams, such as those shown in  Fig.~\ref{fig:linearStability}C, plotting the variable $u$ against $a$ for selected $b$ values.\\

At $b=0.5$ (Fig.~\ref{fig:linearStability}C.2), a single fixed point that
varies monotonically with $a$ is present. This point undergoes Hopf
bifurcations at H$_+$ and H$_-$, as indicated in
Fig.~\ref{fig:linearStability}B, leading to (in)stability of the fixed point as shown in
Fig.~\ref{fig:linearStability}C.2, where solid and dashed lines denote stable and
unstable solutions, respectively. Unstable fixed points evolve into relaxation
oscillations as illustrated in  Fig.~\ref{fig:linearStability}A.1, with limit
cycles forming and disappearing at H$_+$ and H$_-$. Through
path continuation techniques, we can trace these cycles and assess their
stability~\cite{doedel_numerical_1991}. In this parameter regime, oscillations
are stable and exhibit amplitude variations as seen in
Fig.~\ref{fig:linearStability}C.2.
Close to H$_+$ and H$_-$ the oscillations undergo a
\textit{canard explosion} characterized by the abrupt increase of the cycle
amplitude for small changes of $a$~\cite{bold_forced_2003}.\\

As $b$ increases, a cusp bifurcation at $(a,b)=(0,2)$ gives rise to
three new equilibria: $u_{\text{FP}}^b$, $u_{\text{FP}}^m$, and $u_{\text{FP}}^t$.
Fig.~\ref{fig:linearStability}C.3 for $b=2$ illustrates this, where
H$_+$ and H$_-$, previously associated with opposite
$a$ values, now occur close to SN$^r$ and SN$^l$,
respectively.  This diagram corresponds to the most right vertical dashed line
shown in the close-up view of  Fig.~\ref{fig:linearStability}B. Here the top and
bottom equilibrium branches ($u_{\text{FP}}^{t,b}$) are stable nodes and therefore yield
a bistable regime. The former stable limit cycle has vanished due to a
\textit{Saddle-Node bifurcation of Limit Cycles} (SNLC) occurring close to
H$_+$ and H$_-$ (not shown here). At smaller
$\varepsilon$ values, the SNLC occurs near H$_+$ and
H$_-$, these bifurcations and the \textit{Bautin} points from which they arise are thoroughly explored in Rocsoreanu's work~\cite{5}. Additionally,
an unstable limit cycle, emerging from the inverted H$_+$ and
H$_-$, quickly vanishes due to another homoclinic bifurcation occurring
close to the Hopf bifurcation. For a deeper analysis of these phenomena at
larger $\varepsilon$ values, Rocsoreanu \textit{et al.}'s book offers comprehensive
insights~\cite{5}. Adjacent to this, type II excitability emerges, akin to that
shown in  Fig.~\ref{fig:linearStability}A.2, characterized by its proximity to
Hopf/SNLC bifurcations, facilitated by the time scale separation and the
influence of the former limit cycle in the phase space. As mentioned in the
introduction, this type of excitability is characterized by the absence of a
clearly defined threshold.\\

For $b<0$, the system's bifurcation scenario changes significantly. An
illustration for $b=-0.5$ is provided in  Fig.~\ref{fig:linearStability}C.1,
where three fixed points persist, albeit with an opposite slope compared to the
scenario in  Fig.~\ref{fig:linearStability}C.3. Unlike the bistable regime observed
for $b>1$, the top and bottom fixed points ($u_{\text{FP}}^{t,b}$) become
unstable saddle points. Additionally, Hopf bifurcations that initially occurred
at $u_{\text{FP}}^{t,b}$ now emerge in the middle branch $u_{\text{FP}}^m$, rendering it
stable between SN$^l$ and H$_-$, as well as between
H$_+$ and SN$^r$. The relaxation oscillations stemming from
H$_{\pm}$ also experience a canard explosion. Further decreasing
$b$ leads to a homoclinic bifurcation (Hom. in
Fig.~\ref{fig:linearStability}B), where the oscillation period diverges,
potentially exhibiting type-I
excitability~\cite{izhikevich_dynamical_2007,yelo-sarrion_neuron-like_2022}.
This aspect remains largely unexplored within the FHN model.\\

The $(a,b)$-phase diagram can be divided into six principal dynamical regimes (I-VI), as depicted in  Fig.~\ref{fig:linearStability}B:

\begin{itemize}
    \item[I:] \textbf{Monostable regime with a single fixed point.} Defined by
     a single stable node for {$b>0$}, this area splits into
  sectors I.1 and I.2 based on $a$'s positive and
  negative values, respectively. Both sectors exhibit type I
  excitability near the Hopf bifurcation, particularly in the
  presence of canard explosions~\cite{izhikevich_dynamical_2007}.\\

    \item[II:] \textbf{Bistable regime.} Situated between the saddle-node bifurcations SN{$^{l,r}$}, this zone's boundaries are demarcated by H$_+$ for $a>0$ and H$_-$ for $a<0$, featuring three equilibria: two stable nodes ($u_{\text{FP}}^{b,t}$) and one saddle point ($u_{\text{FP}}^{m}$).\\

    \item[III:] \textbf{Oscillatory regime.} This area is known for relaxation oscillations and is further divided into three subregions. In III.1 (for {$b<0$}), it is confined by Hopf bifurcations H$_{\pm}$ and the homoclinic bifurcation Hom.; in III.2 (for $b>0$), oscillations are bounded by H$_{+,-}$ and cease after crossing the SNLC, which occurs near the Hopf bifurcation before the intersection of H$_+$ and H$_-$. Canard explosions are observable near the Hopf bifurcations in each subregion.\\

    \item[IV:] \textbf{Monostable regime within a tri-valued region.} For {$b<0$}, this category includes two monostable regimes where a single stable node ($u_{\text{FP}}^m$) coexists with two saddle points ($u_{\text{FP}}^{b,t}$). These settings are located between SN$^l$ and H$_+$ (IV.1) and between H$_-$ and SN$^r$ (IV.2). For $b>0$, a shift from bistable to monostable occurs due to H$_+$ (for $a>0$) and H$_-$ (for $a<0$), leading to thin monostable strips flanked by H$_+$ and SN$^r$ (for $a<0$) or by H$_-$ and SN$^l$ (for $a>0$). \\

    \item[V:] \textbf{Divergent dynamics regime.} Characterized by divergent
     dynamics due to the lack of an attractor, this regime is prominent
  for {$b<0$} and includes three main divergent sectors:
  V.1, below SN$^l$, with only the saddle point
  $u_{\text{FP}}^b$; V.2, above SN$^r$, with just the saddle
  point $u_{\text{FP}}^t$; and V.3, within the bounds of Hom.,
  H$_+$, and H$_-$, where the unstable node
  $u_{\text{FP}}^m$ coexists with saddles $u_{\text{FP}}^{b,t}$. Proximity to
  the homoclinic bifurcation may induce type I
  excitability~\cite{izhikevich_dynamical_2007}.
\end{itemize}

\begin{figure}
\includegraphics[width=\textwidth]{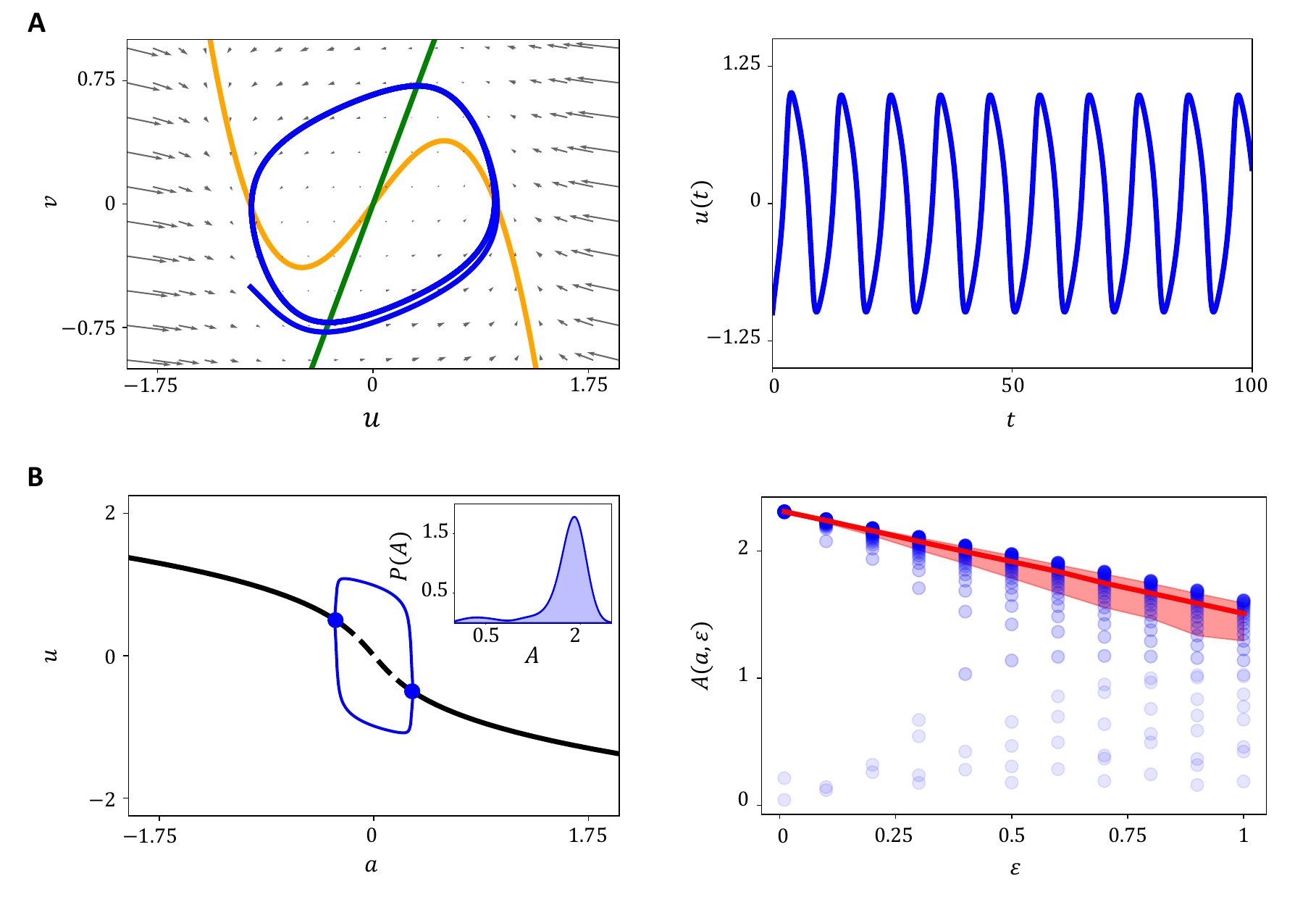}
\caption{\textbf{Time scale separation influences the shape and amplitude of the limit cycle.} \textbf{A}. Effect of $\varepsilon$ on the dynamical behavior in phase space and time series for parameters $(a,b)=(0,0.5)$. Setting $\varepsilon=0.5$ leads to less time scale separation and more sinusoidal oscillations. \textbf{B}. Relaxation oscillations disappear for larger values of $a$ and $\varepsilon$. The oscillation amplitude $A(a,\varepsilon)$ becomes dependent on the $a$ parameter, evidenced by the density plot in the left panel and the dispersion of scatter points in the right panel. Furthermore, an inverse relationship between amplitude and $\varepsilon$ is noted, with a decreasing trend in amplitude as $\varepsilon$ increases, indicated by the median and the 50\% interquartile range values.\label{fig:epsilonDependence}}
\end{figure}

Previously introduced yet not fully explained phenomena, including canard explosions, homoclinic bifurcations (Hom.), and saddle-node bifurcations of limit cycles (SNLC), warrant further exploration to better understand their roles in the dynamics observed.\\

A \textit{canard explosion} refers to a quick increase in the limit cycle's
amplitude as the control parameter increases, generating a cycle that occupies
a substantial portion of the phase space, instead of a small amplitude
oscillation~\cite{Canards1,Canards2}. This phenomenon, illustrated in Figs.
\ref{fig:linearStability}A.1, \ref{fig:linearStability}C.1, and
\ref{fig:linearStability}C.2, is also a characteristic of the Hodgkin-Huxley
model~\cite{HodgkinHuxleyCanards1,HodgkinHuxleyCanards2,HodgkinHuxleyCanards3,25}
and results from the separation of time scales, a property preserved in the FHN
model~\cite{65,13,25,5}. The resilience of canard cycle amplitudes to parameter
perturbations underscores their significance, with the presence of canards
signaling type II excitability -- a persistence of trajectory behavior after a
limit cycle disappears.\\

A \textit{homoclinic bifurcation} represents a global bifurcation where a limit
cycle vanishes upon colliding with a saddle point. This dynamic, akin to
behaviors seen in the Hodgkin-Huxley
model~\cite{HodgkinHuxleyHomoclinic1,HodgkinHuxleyHomoclinic2}, complicates the
identification of some limit cycles within the tri-valued region, especially
for large time-scale separations (small $\varepsilon$), as the homoclinic
bifurcation occurs faster due to the fast increase in amplitude.\\

A \textit{saddle-node bifurcation of limit cycles} occurs when an unstable and a stable limit cycle
collide and annihilate each other, a complexity also found in the
Hodgkin-Huxley model~\cite{HH_SaddleNodeCycles}.  The existence of a
SNLC suggests an unstable cycle nestled within a stable limit
cycle. Therefore, in the tri-valued domain, initializing the system away from
the fixed points is crucial to converge towards the limit cycle. Failure to do
so results in convergence to the stable fixed points, leading to bistable
behavior.\\

This analysis offers an overview of the key dynamical regimes and
behaviors that one could expect in the local FHN model. Given the complex
nature of these dynamics, a thorough exploration extends beyond the scope of
this summary. For readers seeking an in-depth understanding, we suggest
consulting Rocsoreanu's work~\cite{5}, specifically chapter 5, where these
regimes are explored in greater detail.

\subsection{Time scale separation}
\label{timeScaleSeparation}

Finally, we address the impact of the parameter $\varepsilon$ on the FHN model. Unlike $a$ and $b$, which influence the nullclines directly, $\varepsilon$ modifies the time scale separation, thereby adjusting the speed of dynamics in the variable $v$. This effect is evident in the vector field displayed in the first panel of  Fig.~\ref{fig:epsilonDependence}A. It is important to note that in systems with pronounced time scale separation, the oscillation period is predominantly determined by the slower segments. As the dynamics in these segments accelerate, the oscillation period diminishes (Fig.~\ref{fig:epsilonDependence}A, right panel), leading to the disappearance of relaxation oscillations when the separation is minimal. Therefore, time scale separation, alongside multistability, plays a crucial role in enabling excitability and relaxation oscillations, which are central to the FHN model's utility. Additionally, the transition to relaxation-like oscillations is significantly influenced by the degree of time scale separation. As depicted in  Fig.~\ref{fig:epsilonDependence}B, increasing $\varepsilon$ -- which reduces time scale separation -- results in oscillation amplitudes that vary with $a$, moving away from the constant amplitude characteristic of relaxation oscillations.

\section{The diffusively coupled {FitzHugh}-Nagumo system}
\label{diffusive}
When developing the FHN model~\cite{2}, Nagumo \textit{et al.}  extended the original
equation by incorporating diffusive coupling terms, facilitating the
transmission of excitation pulses along the axon. This was
practically demonstrated through the construction of interconnected FHN
electronic circuits, which effectively mimicked pulse propagation, as depicted in
Fig.~\ref{fig:FHNmodel}C. They also explored the thresholds at which these pulses
get excited.\\

Originally conceived for neuroscience applications, the FHN model~\cite{1} and
its spatial extension~\cite{2} have been used beyond their initial scope,
finding relevance across various disciplines. In cardiology, the model is
celebrated for its ability to simulate spiral wave
dynamics~\cite{14,66,67,71,79,80,84,93,102,105,124,130,159,166,177,180}, while
more generally in biology, it is recognized for generating traveling waves that
effectively convey information~\cite{64,70,74,148,149,160,173,174,175,189}. The
model's utility extends to chemistry and computational
sciences~\cite{6,52,61,82,167,171}, where it aids in understanding complex
systems. Beyond its interdisciplinary applications, the FHN model has been
instrumental in investigating mathematical
properties~\cite{30,31,32,33,34,35,36,37,38,39,40,41,42,43,44,45,46,47,48,49,50,51,53,54,55,56,57,59,60,68,69,72,73,75,76,83,85,86,87,88,89,90,91,99,101,103,109,129,150,151,152,153,154,157,171,181,183,184,186,188},
establishing it as a fundamental model in theoretical studies. This section
aims to provide a theoretical exploration into the existence and linear
stability of stationary homogeneous solutions and the emergence of spatially
structured dynamics such as traveling waves and extended patterns. For in-depth
mathematical treatments, readers are directed to~\cite{STspatial}.

\subsection{Linear stability analysis of stationary homogeneous solutions}

Introducing spatial coupling into the analysis does not change the stationary spatially homogeneous solutions or their bifurcation diagrams. However, it requires the consideration of spatially extended perturbations in the stability analysis. Spatial feedback, mediated by diffusion and nonlinearity, often becomes significant near bifurcation points where various effects balance out. Consequently, destabilization of the homogeneous solution may give rise to complex spatial patterns. To elucidate the nature of these instabilities, we examine small temporal and spatial perturbations around the homogeneous states, expressed as $(u,v)=(u_{\text{HS}},v_{\text{HS}})+\epsilon(\xi_u,\xi_v)e^{i{\mathbf{k}}\cdot{\mathbf{x}}+\sigma t}+c.c$., where $\epsilon\ll1$. This leads to an eigenvalue problem:
\begin{equation}
    (J-\sigma I_{2\times2})
\begin{pmatrix}
        \xi_u\\
        \xi_v
\end{pmatrix}
    =
\begin{pmatrix}
        0\\
        0
\end{pmatrix}
,
\end{equation}
with the Jacobian matrix $J$ defined as:
\begin{equation}
    J\equiv  
\begin{pmatrix}
        1-D_uk^2 -3u_{\text{HS}}^2  & -1\\
        \varepsilon & -D_vk^2-\varepsilon b 
\end{pmatrix}
.
\end{equation}

The perturbations' growth rate, determined by the eigenvalues, now varies with the wavenumber $k$, as shown by:
\begin{equation}\begin{array}{ll}
    {\rm Det}[J](k) = D_uD_vk^4+(3D_vu_{\text{HS}}^2-D_v+D_u\varepsilon b)k^2+\varepsilon(3bu_{\text{HS}}^2-b+1), \vspace{3pt}\\ 

    {\rm Tr}[J](k) = -(D_u+D_v)k^2+1-3u_{\text{HS}}^2-\varepsilon b.
\end{array}
\label{EqDiffusiveTraceDeterminant}
\end{equation}
Setting $k=0$ recovers the instabilities found in the non-spatial
system, such as Hopf and saddle-node bifurcations. The dispersion relation
$\sigma(k)$ reveals that a variety of coherent structures, including traveling
waves and stationary spatial patterns, emerge from distinct instabilities (for an in-depth discussion, see
Refs.~\cite{36,51,72,83,85,90,99,101,149,150,174,183}). The role of
nonlinearity is critical in determining the existence and stability of these
predicted structures. Our focus here is on the most prominent spatially
coherent structures supported by the FHN model.

\subsection{Turing patterns in one spatial dimension}

\begin{figure}
\includegraphics[width=\textwidth]{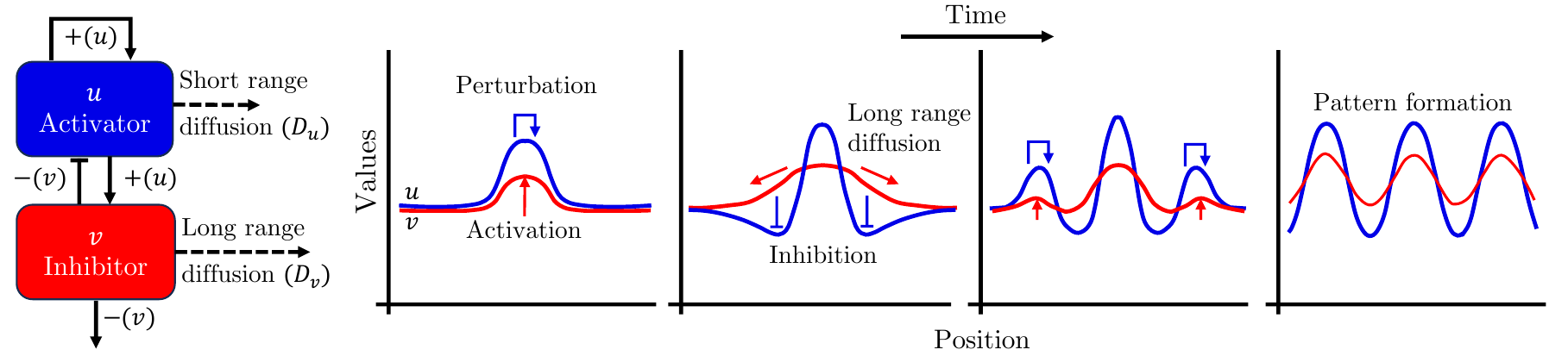}
\caption{\textbf{Schematic illustration of pattern formation in the FHN model.} The interplay of components that enable Turing patterns through scale-dependent feedback is depicted (left). Initially, a small perturbation amplifies via autocatalysis, concurrently triggering the inhibitor. The inhibitor's rapid diffusion, stemming from differential diffusion rates, subsequently suppresses the activator, indirectly inhibiting itself. This spatially propagating interaction can lead to a regular pattern.\label{fig:EssentialIngredients}}
\end{figure}

Spatial patterns, resulting from self-organizing processes, are widely observed
in nature and have been a subject of interest across various scientific fields.
In developmental biology, for example, such patterns provide a mechanism for
breaking the symmetry in the initially homogeneous tissue of an organism,
guiding the development of complex structures~\cite{kondo2010}. Similarly, in
ecology, the self-organization of vegetation patterns plays a crucial role in
ecosystem resilience to environmental fluctuations~\cite{rietkerk2021}. Despite
the diverse manifestations of self-organization, the underlying mathematical
frameworks share a foundational theory, first introduced by Alan Turing in
1952~\cite{Turing}. Turing's theory of morphogenesis highlighted the essential
elements for spatial self-organization within reaction-diffusion systems like
the FHN model. In these systems, an activator substance ($u$) promotes
the production of its inhibitor ($v$), which diffuses more rapidly and
suppresses the activator in adjacent areas, establishing a feedback mechanism
that generates regular spatial patterns, as depicted in
Fig.~\ref{fig:EssentialIngredients}. The FHN model, while primarily explored from a
theoretical perspective, serves as a valuable tool for studying pattern
formation, particularly given its oscillatory dynamics that contribute to the
complexity of the observed patterns (see  \ref{SS:pattern2D}).\\

\begin{figure}
\includegraphics[width=\textwidth]{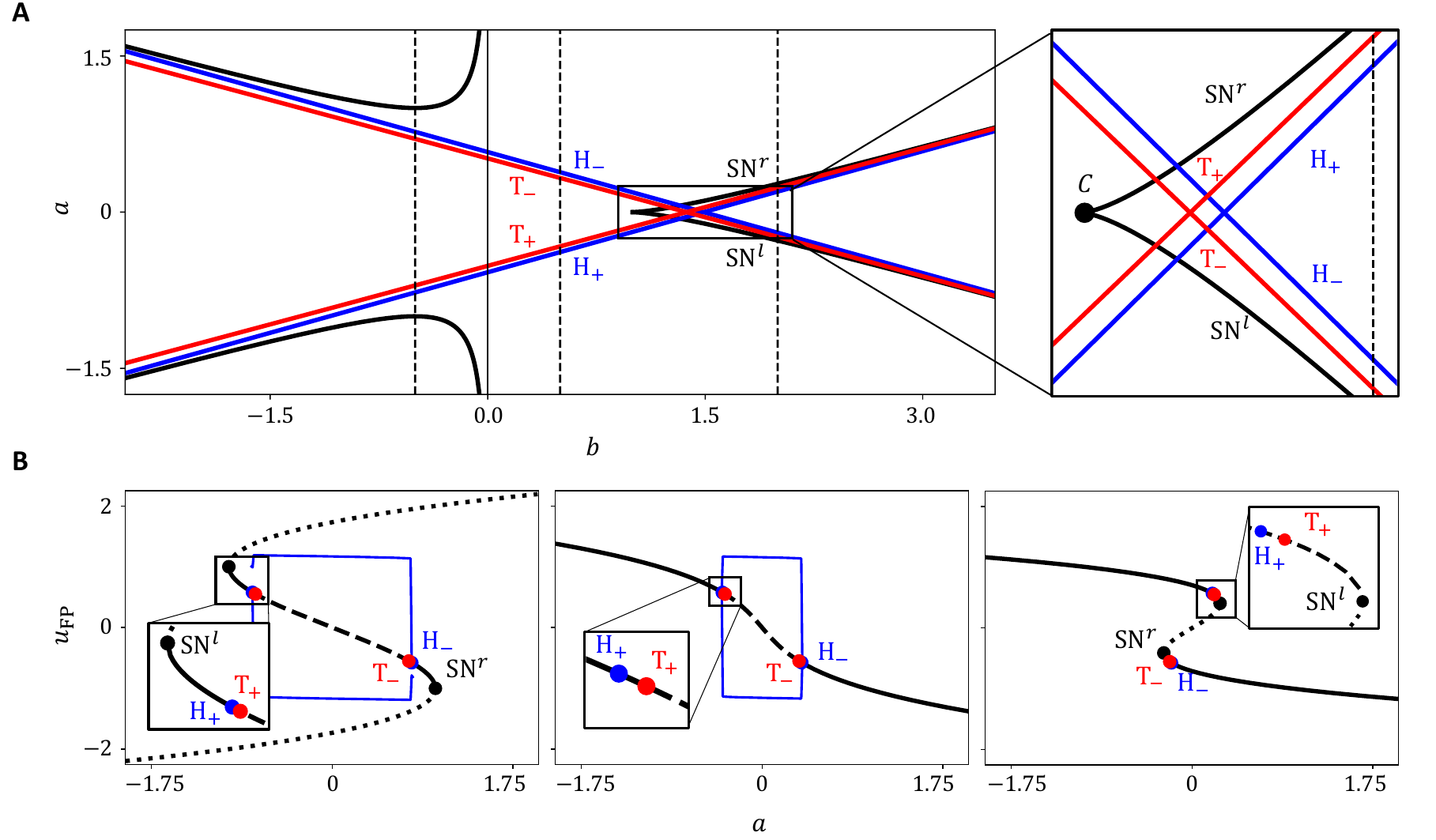}
\caption{\textbf{Mapping of the Turing bifurcation in} $(a,b)$ \textbf{space}. \textbf{A}. The phase diagram, highlighting Turing bifurcations T$_\pm$, Hopf bifurcations H$_\pm$, and saddle-node bifurcation, with a zoomed-in view of the cusp (C) within the black rectangle. \textbf{B}. Bifurcation diagrams for selected $b$ values of $-0.5, 0.5$, and $2$, as marked by vertical dashed lines in A, with parameters set to $D_u=1$, $D_v=1$, and $\varepsilon=0.01$.\label{fig:SpatialBifurcations1}}
\end{figure}

\begin{figure}
\includegraphics[width=\textwidth]{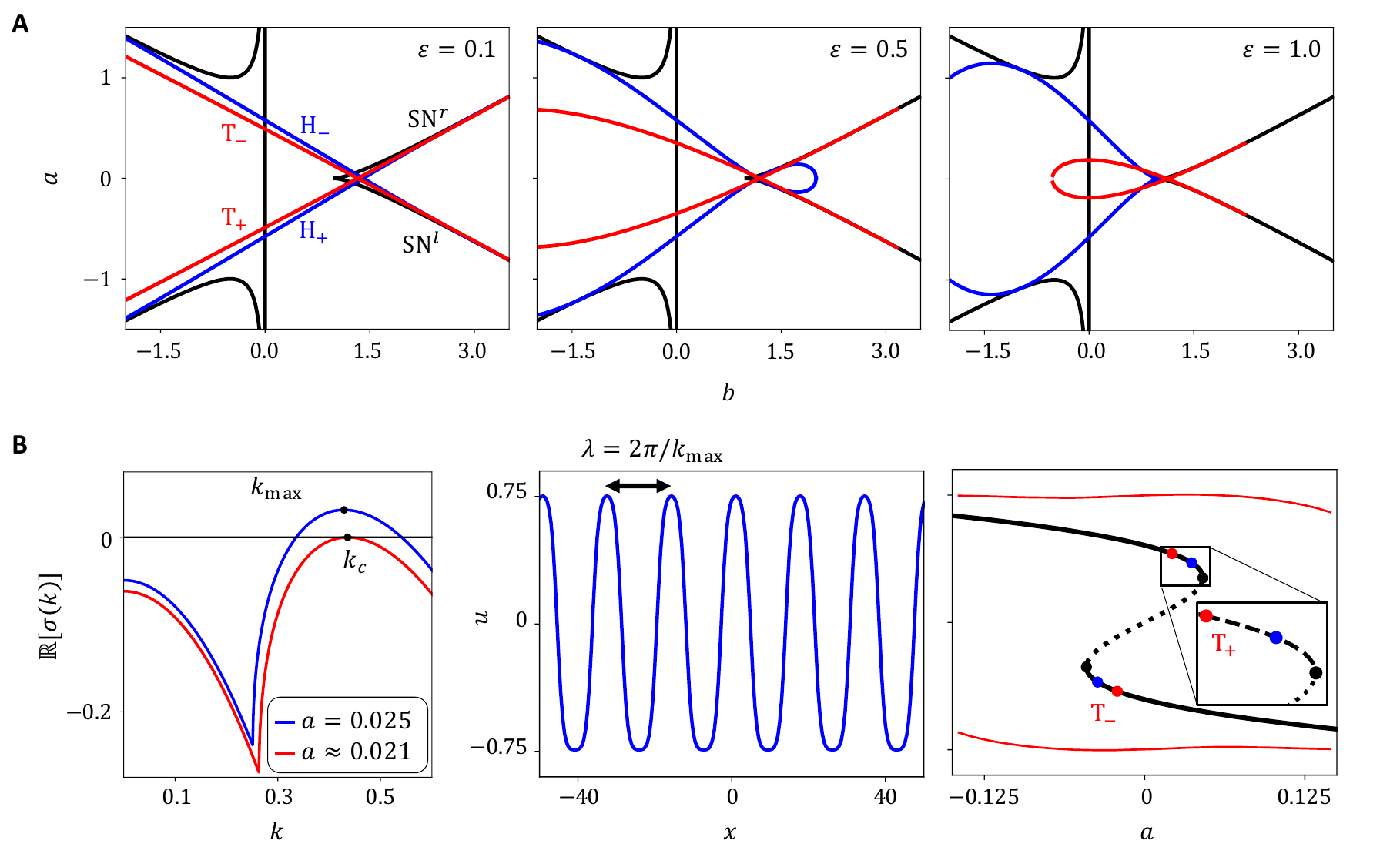}
\caption{\textbf{Exploring pattern formation with reduced time-scale separation.}  \textbf{A}. The $(a,b)$-phase diagram's transformation is illustrated, highlighting the shifts in Hopf and Turing bifurcations with varying $\varepsilon$, given $D_u=1$ and $D_v=5$.  \textbf{B}.  Turing pattern emergence at $a=0.025$, $b=1.26$, and $\varepsilon=0.5$, as indicated by the vertical dashed line in A. The growth rate as a function of wavenumber $k$ is shown for $a=a_T$ (left), followed by the resulting Turing pattern at the most unstable wavelength (center), and a bifurcation diagram tracing the pattern's maxima and minima across varying $a$ values (right), utilizing the same parameters as in  Fig.~\ref{fig:SpatialBifurcations1}.\label{fig:SpatialBifurcations}}
\end{figure}

Pattern formation often emerges at the juncture of competing effects, typically
near other bifurcations of a system's homogeneous state where specific
eigenvalues converge towards zero. In the FHN model, Turing
instabilities~\cite{Turing}, which lead to pattern formation, appear close to
the saddle-node bifurcations, illustrated in Figs.~\ref{fig:SpatialBifurcations1} and \ref{fig:SpatialBifurcations}.
A Turing instability, sometimes also called a modulation instability, marks the transition where a uniformly stable state becomes unstable to perturbations of a specific wavelength $k_c$. This critical point is identifiable in the dispersion relation when the growth rate of mode $k_c$ turns positive (Fig.~\ref{fig:SpatialBifurcations}B), signaling the temporal increase of this mode's amplitude and the onset of spatially periodic Turing patterns (Fig.~\ref{fig:EssentialIngredients}). The bifurcation conditions, ${\rm Det}[J](k)=0$ and $\partial_k {\rm Det}[J](k)=0$, must be simultaneously met at a non-zero wavenumber $k=k_c$, leading to the critical wavelength and defining the instability's location:
\begin{equation}
\label{EqTuring}
k_c=\pm\sqrt{\dfrac{-3D_v{u_{\text{HS}}}^2+D_v-D_u\varepsilon b}{2D_uD_v}},
\end{equation}
\begin{equation}
u_\text{T}=\pm\sqrt{\dfrac{1}{3D_v}\left(D_v+D_u \varepsilon b\pm 2\sqrt{D_uD_v\varepsilon}\right)},
\end{equation}
where the criterion $D_v>3D_vu_{\text{HS}}^2+D_u\varepsilon b$ must be satisfied.\\

The criterion for instability requires that the inhibitor's diffusion rate
surpasses that of the activator ($D_v>D_u$), aligning with Murray's findings
for a general system~\cite{63}. The Turing bifurcation's dependence on
parameters $a$ and $b$ mirrors the Hopf bifurcation's
parameter dependence, as depicted in  Fig.~\ref{fig:SpatialBifurcations1}A. These
Hopf and saddle-node bifurcations correspond to the spatially uniform state
instabilities shown in  Fig.~\ref{fig:linearStability}B. As $b$ increases
significantly, the Turing and Hopf instabilities converge towards the saddle-node (SN)
bifurcations (black lines) of the uniform state, indicating the homogeneous
stationary state's stabilization (or destabilization) with increasing (or
decreasing) $b$.\\

Particularly intriguing dynamics emerge near the cusp bifurcation $C$, where a region exhibiting coexisting bifurcations transitions from bistable to monostable regimes, highlighted in the inset of  Fig.~\ref{fig:SpatialBifurcations1}A. Different $b$ values shift the system from monostable to bistable regimes, influencing the bifurcations' positions. The vertical dashed lines in  Fig.~\ref{fig:SpatialBifurcations1}A mark the bifurcation diagrams in  Fig.~\ref{fig:SpatialBifurcations1}B for $b=-0.5$, $b=0.5$, and $b=2$, showing that the Turing instability, occurs after the uniform states become Hopf unstable. At this value of $\varepsilon$, static Turing patterns are absent due to the faster growth of the homogeneous mode compared to the nonzero wave number mode.\\

To achieve robust pattern formation, increasing $\varepsilon$ is necessary.  Fig.~\ref{fig:SpatialBifurcations}A illustrates how the $(a,b)$-phase diagram transforms with varying $\varepsilon$, where the oscillatory region between the Hopf lines H narrows, and at $\varepsilon=1$, it confines to a lobe touching $C$ at $(a,b)=(0,1)$. It is noteworthy from Eq.~\ref{EqTuring} that similar effects occur when increasing $D_u/D_v$ instead of $\varepsilon$, since in both the dependence goes as $\propto \pm\sqrt{x\pm\sqrt{x}}$. Nonetheless, it is crucial to remember that pattern formation is contingent on $D_v>D_u$.\\

With an increased value of $\varepsilon$, the difference in time scales between
the variables diminishes, enhancing scale-dependent feedback, resulting in the
Turing instability manifesting prior to the Hopf bifurcation. By fixing
$b$ at 1.26 and setting $\varepsilon$ to 0.5 (see the vertical dashed line in Fig.~\ref{fig:SpatialBifurcations}A), the bifurcation diagram reveals the emergence of
spatially periodic patterns due to a Turing instability, illustrated in
Fig.~\ref{fig:SpatialBifurcations}B. Near the Turing bifurcation point, the growth
rate for modulated perturbations on the uniform state turns positive, signaling
the start of instability, as indicated in  Fig.~\ref{fig:SpatialBifurcations}B
(left). Beyond this threshold, the dominant wavenumber $k_u$ initiates
the formation of regular patterns, shown in  Fig.~\ref{fig:SpatialBifurcations}B
(center). The evolution of pattern maxima and minima with $a$,
obtained from direct numerical simulations, is plotted with a continuous line
in  Fig.~\ref{fig:SpatialBifurcations}B (right). Given these parameters, the
pattern coexists with homogeneous solutions across a broad range in the parameter $a$,
suggesting a potential subcritical emergence from the Turing instability,
warranting further examination for confirmation. Unlike other pattern-forming
systems, the symmetry in the FHN model leads to identical stationary patterns
beyond T$_{\pm}$. This symmetry implies that mode $k_c$ may
bifurcate subcritically from T$_+$, stabilize upon folding at lower
$a$ values, and persist until nearing a second fold at a symmetrical
$a$ value, then transitioning to the second Turing T$_-$, akin
to other models~\cite{thiele2019,parra-rivas_organization_2023}. To our
knowledge, this bifurcation analysis of 1D patterns in the FHN model is new and
unreported in existing literature, highlighting the need for more in-depth
exploration of spatial structure bifurcations, particularly the stability of
modes bifurcating after the Turing bifurcation far from the onset or the
associated localized states.\\

Furthermore, pattern instabilities like the Eckhaus instability, characterized
by the destabilization of high-wavenumber periodic solutions by long-wavelength
perturbations, might also occur~\cite{eckhaus_book}. This secondary instability
can cause primary pattern distortion or fragmentation. Although analytically
tractable in simpler models like the Ginzburg-Landau equation~\cite{eckhaus},
it has also been investigated in the FHN context~\cite{72}, examining phase
dynamics of near-stationary patterns across various parameter regimes, even
well beyond the Turing instability onset~\cite{hagberg_phase_2000}. Subsequent
sections will provide additional examples of these instabilities and Turing
patterns, especially within two-dimensional contexts, where they are
predominantly studied.

\subsection{Fronts and localized structures}
\label{localized_structures}

Front solutions play a pivotal role in bistable and excitable
reaction-diffusion systems, with the FHN model being a classic example. While
traditionally not a focus within neuroscience, the study of front solutions has
garnered attention for their implications in pattern formation and their
ability to transmit information effectively. This has made them particularly
relevant in fields like cardiology~\cite{79,84,180}, other biological
systems~\cite{64,74,173,175,189}, and even geology~\cite{160}, despite the FHN
model's relative simplicity limiting most front-related studies to mathematical
explorations aimed at understanding general front
properties~\cite{30,32,34,35,36,39,42,44,47,56,73,84,87,90,91,101,150,152,154,181,183}.\\

\begin{figure}
\includegraphics[width=\textwidth]{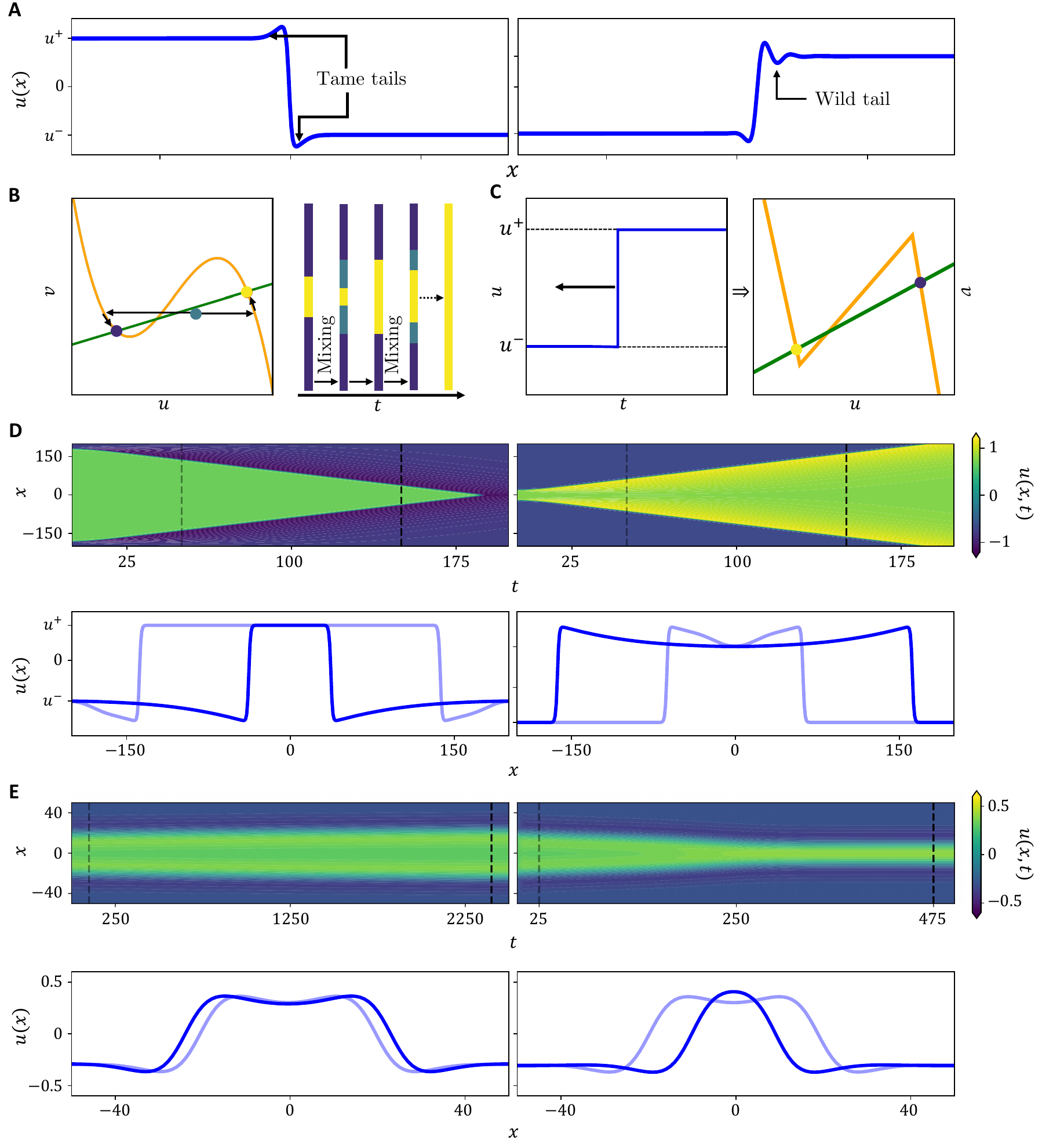}
\caption{\textbf{Dynamics of localized states through the lens of various front behaviors.} \textbf{A}. Two contrasting front types - a ``tame'' front (left) and a ``wild'' front (right), with the tame front parameters set at $(D_u, D_v, a, b, \varepsilon) = (1, 1, 0, 2, 0.01)$ and the wild front at $(0.49, 1, 0.01, 1.1, 1)$. \textbf{B}. Sketch of how a traveling front can be created. \textbf{C}. The commonly used piecewise linear approximation leverages the separation of time scales to analytically predict front properties. \textbf{D}. Kymographs and their corresponding profiles for traveling tame fronts, employing the same parameters as the tame front with added asymmetry by adjusting $a$ to 0.1 (left) and -0.1 (right). \textbf{E}. Formation of localized states via front interactions, with $a$ values set at 10$^{-5}$ (left) and 10$^{-3}$ (right) for the parameters $(D_u, D_v, b, \varepsilon) = (0.49, 1, 1.1, 1)$.\label{fig:fronts}}
\end{figure}

In a one-dimensional space, fronts manifest as \textit{heteroclinic orbits}
that bridge two distinct uniform states, potentially differing in
stability~\cite{homburg_chapter_2010}. These dynamics give rise to two front
types, distinguished by their approach to the uniform state, specifically the
nature of their `tails'. Tails that approach uniformly without oscillations
form what are known as \textit{tame} or flat fronts. Conversely, tails that
exhibit damped oscillations around the uniform state result in \textit{wild
fronts}. Illustrations of both front types within the FHN framework are
provided in~\cite{16,34,39} and are depicted in  Fig.~\ref{fig:fronts}A.\\

Near homogenous states, front behaviors can be linearly approximated by:
\begin{equation}
\label{Approach}
\begin{pmatrix}
    u(x)\\
    v(x)
\end{pmatrix}
-    
\begin{pmatrix}
    u_{\text{HS}}\\
    v_{\text{HS}}
\end{pmatrix}
\propto e^{\lambda x},
\end{equation}
where $\lambda$ represents the spatial eigenvalue derived from $\sigma(-i\lambda)=0$. Tame fronts correspond to purely real $\lambda$ values, whereas complex $\lambda$ values indicate the presence of wild fronts.\\

Like all nonlinear phenomena, front dynamics can exhibit instabilities, leading
to the formation of complex patterns~\cite{56,85}. Specifically, fronts can
become susceptible to transverse modulations, a phenomenon known as transverse
front instability, resulting in the ``fingering'' effect and the subsequent
development of ``labyrinth patterns''~\cite{83}. Additionally, fronts might
experience the nonequilibrium \textit{Ising-Bloch instability}, giving rise to
two counterpropagating fronts~\cite{85}. However, in the context of the FHN
model, the coexistence of a stable planar front with a stable large-amplitude
stripe pattern, a prerequisite for nonlinear front transverse instability, has
yet to be observed~\cite{83}.\\

McKean, in 1969, introduced a simplified approach to the FHN model through a
piecewise approximation (Fig.~\ref{fig:fronts}C), facilitated by the distinct time
scales in the system~\cite{16}. This approximation transforms the cubic
function into a piecewise or Heaviside function, allowing for an analysis of
front formation and propagation within an integro-differential framework of the
FHN model:
\begin{equation}
    \partial_t u=\partial_x^2 u+f(u)-b\int u(t) dt,
\end{equation}
where $f(u) = u(1-u)(u-a)$. This formulation enables the investigation of front
solutions $u = u(x + ct)$, particularly under the condition $b = 0$. McKean's
piecewise approximation has since been applied in various contexts to deduce
conditions for different front behaviors (e.g.~tame vs. wild
fronts)~\cite{32,39,40,152}.\\

Stable fronts result in regular wave propagation, contingent on the presence of
an ``energetically preferred'' state. This is illustrated in scenarios where
regions of high and low activity eventually converge to a preferred state, as
shown in  Fig.~\ref{fig:fronts}B. The concept of ``equivalent uniform'' states, where
two states are related by the $u \rightarrow -u$ transformation and possess equal
energy in systems with definable free energy, leads to static
fronts~\cite{chomaz_absolute_1992}. This equilibrium is attained at the Maxwell
point of the system, identified in the FHN model at $a = a_M \equiv 0$. Front
movement is observed when parameters deviate from this point.\\

The interaction between moving fronts of opposite polarities (i.e.,~$F_{u_b\rightarrow u_t}$ and $F_{u_t\rightarrow u_b}$) can lead to various outcomes, describable by an effective reduced equation:
\begin{equation}
\label{Eq_interaction}
    \frac{dD}{dt}=A e^{{\rm Re}(\lambda)D}\cos({\rm Im}(\lambda) D)+B,
\end{equation}
with $\lambda$ representing the dominant spatial eigenvalue related to the
front, $A$ depending on system parameters, and $B \propto a - a_M$ indicating
the deviation from the Maxwell
point~\cite{coullet_nature_1987,coullet_localized_2002}. This framework
provides a phenomenological understanding of front dynamics and interactions.\\

Returning to the general FHN model,  in scenarios where the wave fronts are
tame, characterized by ${\rm Im}(\lambda)=0$, their attraction or repulsion follows a
monotonous exponential law. This behavior is illustrated in  Fig.~\ref{fig:fronts}.
When the system parameter $a$ is less than $a_M$ (i.e.,~$a<a_M$), the state $u^t$ predominates, causing the wave fronts to
move in opposite directions and eventually dominate the entire domain, as shown
in  Fig.~\ref{fig:fronts}D. Conversely, for $a$ values exceeding
$a_M$ (i.e.,~$a>a_M$), the fronts move towards each other and
ultimately annihilate in a process known as \textit{coarsening}, leading the
system to stabilize at $u_b$~\cite{cahn_free_2004}. \\

In contrast, when the fronts are wild (${ \rm Im}(\lambda)\neq0$), any stationary solution
of Eq.~\ref{Eq_interaction} corresponds to the locking or pinning of two
fronts, resulting in localized states. Near the system's Maxwell point
($a=a_M$, where $B=0$) in an infinite domain,
Eq.~\ref{Eq_interaction} presents countless equilibrium distances with
alternating stability~\cite{parra-rivas_organization_2023}. Deviating from the
Maxwell point ($B\neq0$), the variety of equilibrium distances -- and
consequently, the potential localized states (LSs) -- decreases. The emergence
of two distinct LSs near the Maxwell point is depicted in  Fig.~\ref{fig:fronts}E
for $a=10^{-5}$ (left) and $a=10^{-3}$ (right). In the former scenario, the
fronts lock very fast, resulting in a broad LS, while for $a=10^{-3}$, the
slower convergence of fronts eventually leads to the formation of a narrower,
single-peak LS~\cite{frohoff2023stationary}.

\subsection{Traveling pulses}

Traveling pulse solutions are a hallmark of bistable and excitable
reaction-diffusion systems, a phenomenon first leveraged by Nagumo \textit{et al.} to
model excited pulse propagation along neuronal axons, and subsequently adopted
in numerous studies~\cite{2,40,62}. Beyond neuroscience, these dynamics have
profound implications in cardiology~\cite{14,71,79,159,180}, where the
excitable nature of cardiac tissues is essential for maintaining coordinated
heart rhythms and understanding cardiac dysfunctions. While less emphasized,
the FHN model has also illuminated information propagation mechanisms in
various biological systems~\cite{64,70,173,175,189}, akin to traveling fronts.\\

\begin{figure}
\includegraphics[width=\textwidth]{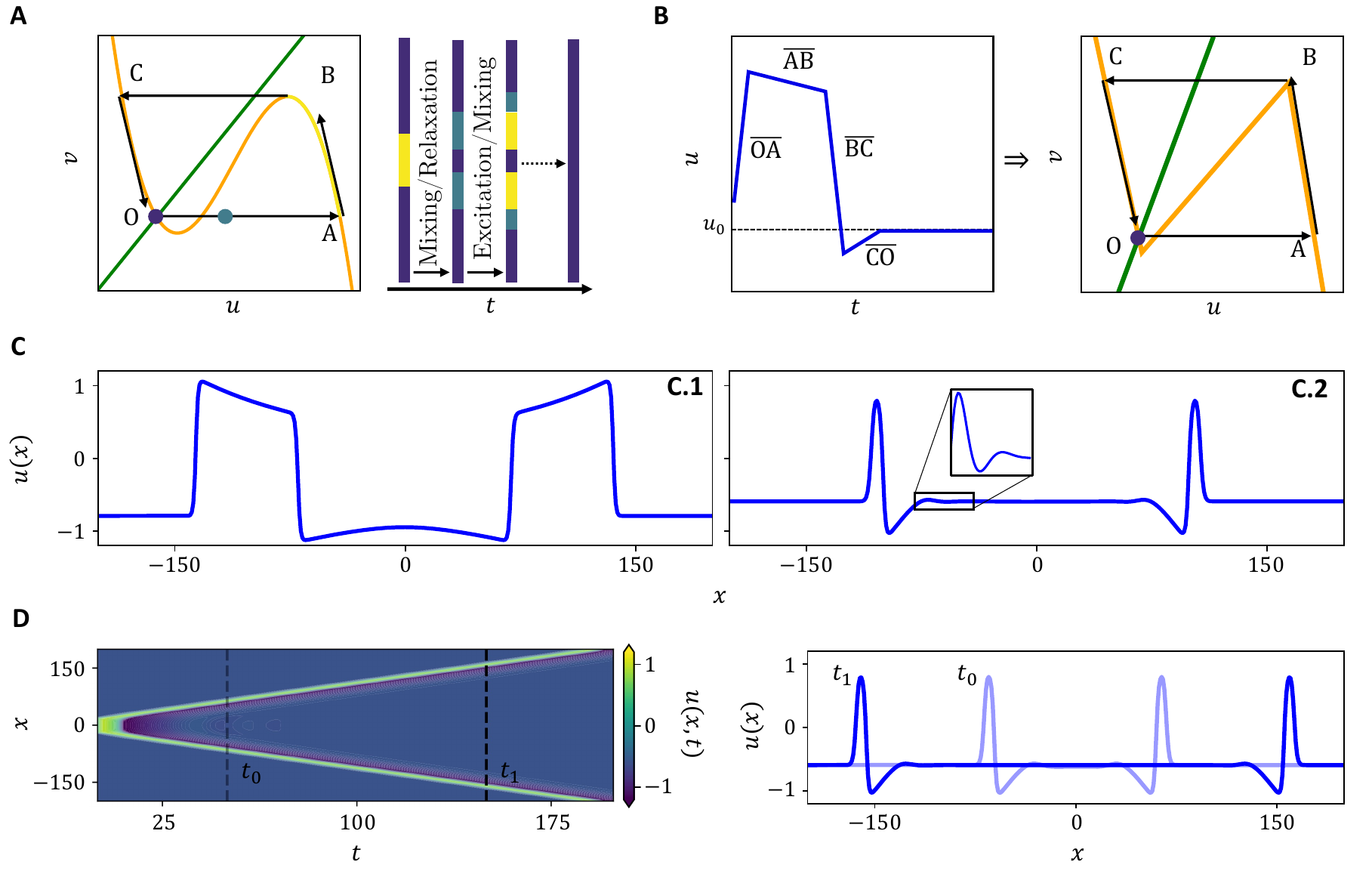}
\caption{\textbf{Excitable traveling pulses in the FHN model}.
    \textbf{A}.  Formation process of an excitable traveling pulse. Initially, the system is disrupted beyond a certain pseudothreshold determined by the unique structure of the folded nullclines. This leads to the diffusion-induced excitation of neighboring regions, causing the pulse to travel through space. Subsequently, areas that were excited earlier trace back along the nullcline, eventually settling into a uniform state.
    \textbf{B}. A simplified representation of the nullclines as piecewise linear functions, an approximation that holds primarily under significant time scale separation.
    \textbf{C}. Two types of pulses: a `tame' pulse (left), generated under the parameters $(D_u,D_v,a,b,\varepsilon) = (1,1,0.2,2,0.01)$, and a `wild' pulse (right), arising from $(1.5,1,0.4,0.5,0.1)$.
    \textbf{D}. Propagation dynamics of the `wild' pulse. The pulse maintains its shape as it moves, shown at two distinct moments, $t_0$ and $t_1$, marked by vertical dashed lines on the kymograph.\label{fig:pulses}}
\end{figure}

Research on excitable pulses within the FHN framework has predominantly
concentrated on type II excitability, the most recognized form. However,
emerging research in ecology, especially regarding vegetation
patterns~\cite{ruiz2023self,zhao2021fairy}, has sparked an exploration of type
I excitable pulses~\cite{moreno2022bifurcation,arinyo2021traveling}, a domain
yet to be explored with the FHN model, despite featuring a homoclinic
bifurcation of the limit cycle as well.\\ 

In this subsection, we focus on the fundamental mathematical characteristics of
traveling pulses, situating them within the broader context of existing FHN
literature and underscoring the model's pivotal role in the mathematical
investigation of pulse
dynamics~\cite{30,31,32,37,41,42,44,45,46,48,49,87,91,99,101,150,151,152,154,157,183}.

In one-dimensional systems, traveling pulses correspond to \textit{homoclinic
orbits} of the spatial system in the comoving reference frame. These orbits
form connections from a stable fixed point back to
itself~\cite{glendinning1994stability}. Illustrated in  Fig.~\ref{fig:pulses}A,
perturbations exceeding a critical threshold from a uniform state induce an
increase in $u$, initiating a substantial excursion that, through
diffusion, activates adjacent regions. The system eventually reverts to the
initial state, descending along the folded nullcline before a sharp reduction
in $u$ leads back to the baseline state along the opposing nullcline.
This propagation mechanism is akin to a trigger wave initiating at the
forefront, succeeded by a phase wave, a process underpinned by the excitable
medium's characteristics, facilitating the formation of traveling
pulses~\cite{tyson1988}.\\

\begin{figure}
\includegraphics[width=\textwidth]{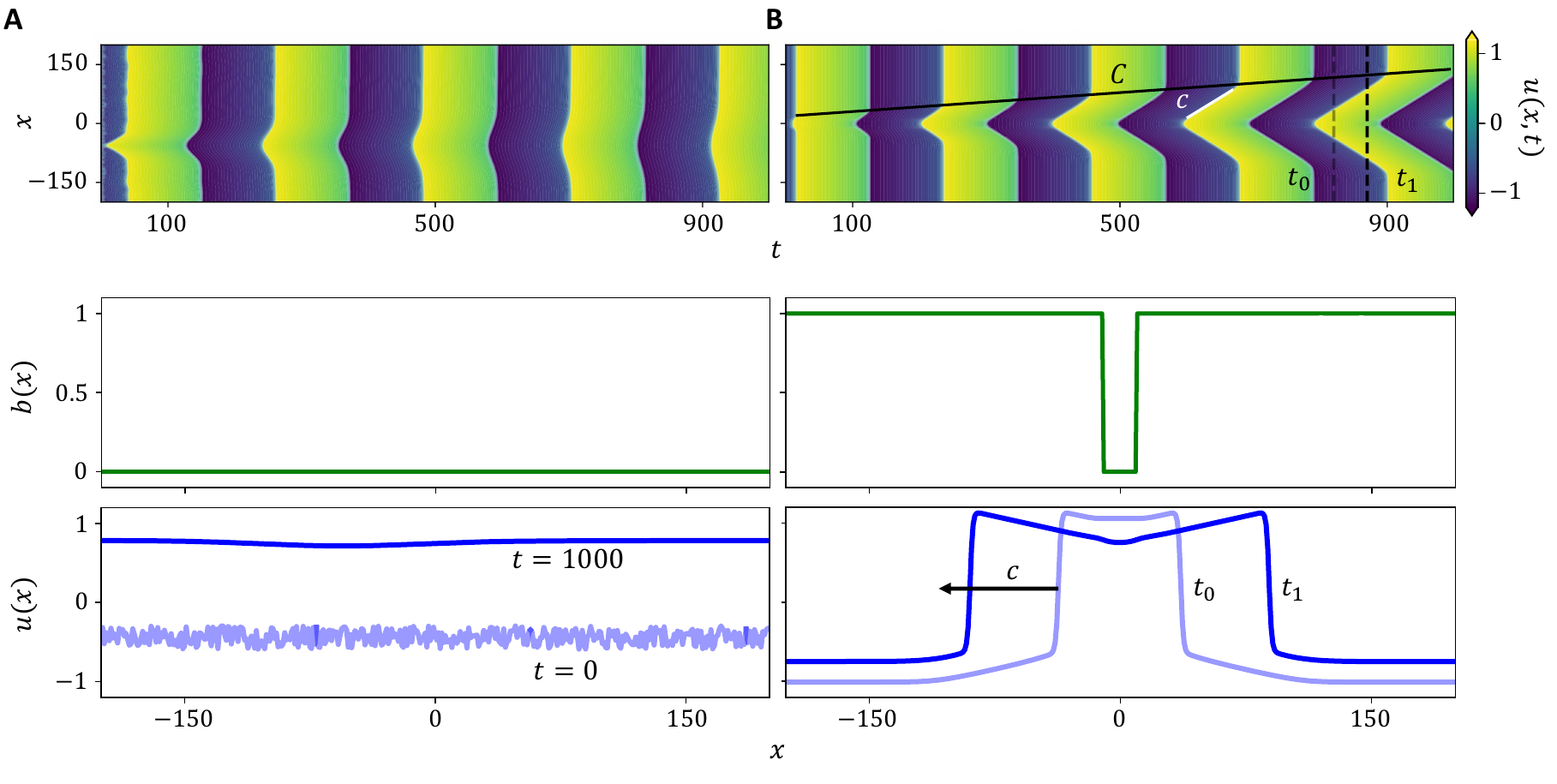}
\caption{\textbf{The role of spatial heterogeneity in generating wave trains.} The simulations are based on parameters $(a, \varepsilon, D_u, D_v) = (0, 0.01, 1, 1)$, with $b$ varying. \textbf{A}. Initial heterogeneities in variables diffuse and eventually dissipate. \textbf{B}. An oscillatory system ($b = 1$) incorporating a central pacemaker zone where $b = 0$ spanning $20$ units. Fast oscillations spread as wave trains at speed $c$, while the influence of the pacemaker expands throughout the system at an envelope speed $C$.\label{fig:Pacemakers}}
\end{figure}

While the generation of excitable traveling pulses is conceptually rather
simple, their investigation is complex. The FHN model is extensively employed
to explore key features of traveling pulses, utilizing its piecewise linear
approximation for simplicity (Fig.~\ref{fig:pulses}B). The insights from
earlier discussions apply here, demonstrating the broader relevance of the
piecewise linear FHN model beyond McKean's work~\cite{16}, with various studies
addressing traveling pulses~\cite{32,40,152,154} and traveling
waves~\cite{39,65} within this framework.\\

Revisiting the categorization from  \ref{localized_structures}, traveling pulses can also be distinguished as tame or wild based on their approach to the uniform state.  Fig.~\ref{fig:pulses}C.1 and C.2 show examples of tame (with parameters $D_u=1$, $D_v=1$, $a=0.2$, $b=2$, $\varepsilon=0.01$) and wild (with parameters $D_u=1.5$, $D_v=1$, $a=0.4$, $b=0.5$, $\varepsilon=0.1$) pulses, respectively. The propagation of a wild pulse is depicted in  Fig.~\ref{fig:pulses}D, where snapshots at times $t_0$ and $t_1$ -- marked by white dashed lines in the kymograph -- illustrate the pulse's consistent shape.\\

From a bifurcation analysis viewpoint, the emergence of excitable pulses is
linked to the destabilization of a uniform state through a finite wavelength
Hopf bifurcation~\cite{41}. Yochelis \textit{et al.}  used a three-dimensional FHN-like
model to demonstrate that near a subcritical finite wavelength Hopf
bifurcation, such states might exhibit homoclinic snaking, yielding multi-pulse
solutions where multiple excitable pulses coexist. For further details on
multi-pulse solutions in the FHN equations, readers are directed to the works
of Kupra~\cite{46} and Hastings~\cite{157}.\\

Up to this point, our discussion has been centered around traveling pulses
emerging from local excitability within the FHN model. Yet, the model also
supports other variants of traveling pulses through different mechanisms. A
notable example is the parity-breaking front bifurcation explored by Elphick
\textit{et al.}~\cite{45}. In their study, they reveal that the FHN model, moving past
the nonequilibrium Ising-Bloch instability, can generate symmetric traveling
pulses that diverge from the previously mentioned excitable pulses.

\subsection{Pacemakers and wave trains}

We now look into the dynamics of oscillatory media coupled via diffusion, a
subject less frequently studied in existing research on the FHN
model~\cite{37,49,53,75,75b}. This aspect is particularly relevant to cellular
and molecular biology, where traveling waves facilitate long-distance system
synchronization and intra- or inter-cellular
communication~\cite{64,Beta_Kruse_review_waves,ditalia2022}.For these
biological processes to be coordinated effectively, it is crucial for the waves to
propagate sufficiently quickly through the medium.\\

In scenarios with initial condition heterogeneities, diffusion tends to
homogenize the system, leading to uniform oscillations across the medium, as
depicted in  Fig.~\ref{fig:Pacemakers}A. This phenomenon highlights the system's
capacity to mitigate initial irregularities through its inherent dynamics,
showcasing an inherent resilience to fluctuations~\cite{64}.

The system can also have heterogeneities in its parameters. Specifically, regions with oscillation frequencies exceeding those of their surroundings can emanate waves, thereby entraining the entire medium, as illustrated in  Fig.~\ref{fig:Pacemakers}B. Such areas are termed pacemakers. The resulting oscillation frequency is a compromise between the fast and slow frequencies of the uncoupled system. The wave's propagation speed (white line on  Fig.~\ref{fig:Pacemakers}B) and the rate at which it influences the surrounding medium, termed the envelope speed (black line on  Fig.~\ref{fig:Pacemakers}B), are important metrics.\\

With a pronounced time scale separation (low $\varepsilon$), wave speeds are
significantly higher than those in systems with less time scale separation,
likely due to front propagation between stable states akin to excited pulse
propagation. They can be similarly studied using a singular perturbation
approach~\cite{tyson1988}. The speeds of wave trains and their envelopes can be
linked through a formula (assuming constant speeds)~\cite{75,75b}:
\begin{equation}
    C=\frac{T_0-T_t}{T_0}c,
\end{equation}
where $C$ represents the envelope speed, $c$ denotes the wave
train speed, and $T_0$ and $T_t$ correspond to the slow uncoupled
period and the resulting period with diffusive coupling, respectively.
Furthermore, speed is closely tied to time scale separation, pacemaker size,
diffusion strength, and initial frequency difference~\cite{75,75b}. Contrary to
intuition, the pacemaker's size and frequency difference distinctively affect
wave and envelope speeds. Larger or higher-frequency pacemakers, deemed
stronger, take over the medium faster via increased envelope speed, albeit at
reduced wave velocities. For smaller time scale separation (high $\varepsilon$),
the oscillations become nearly harmonic, altering wave dynamics from linear
front-driven to a sublinear spread, better analyzed through phase-reduction
methods~\cite{75b}. \\

\begin{figure}
\includegraphics[width=\textwidth]{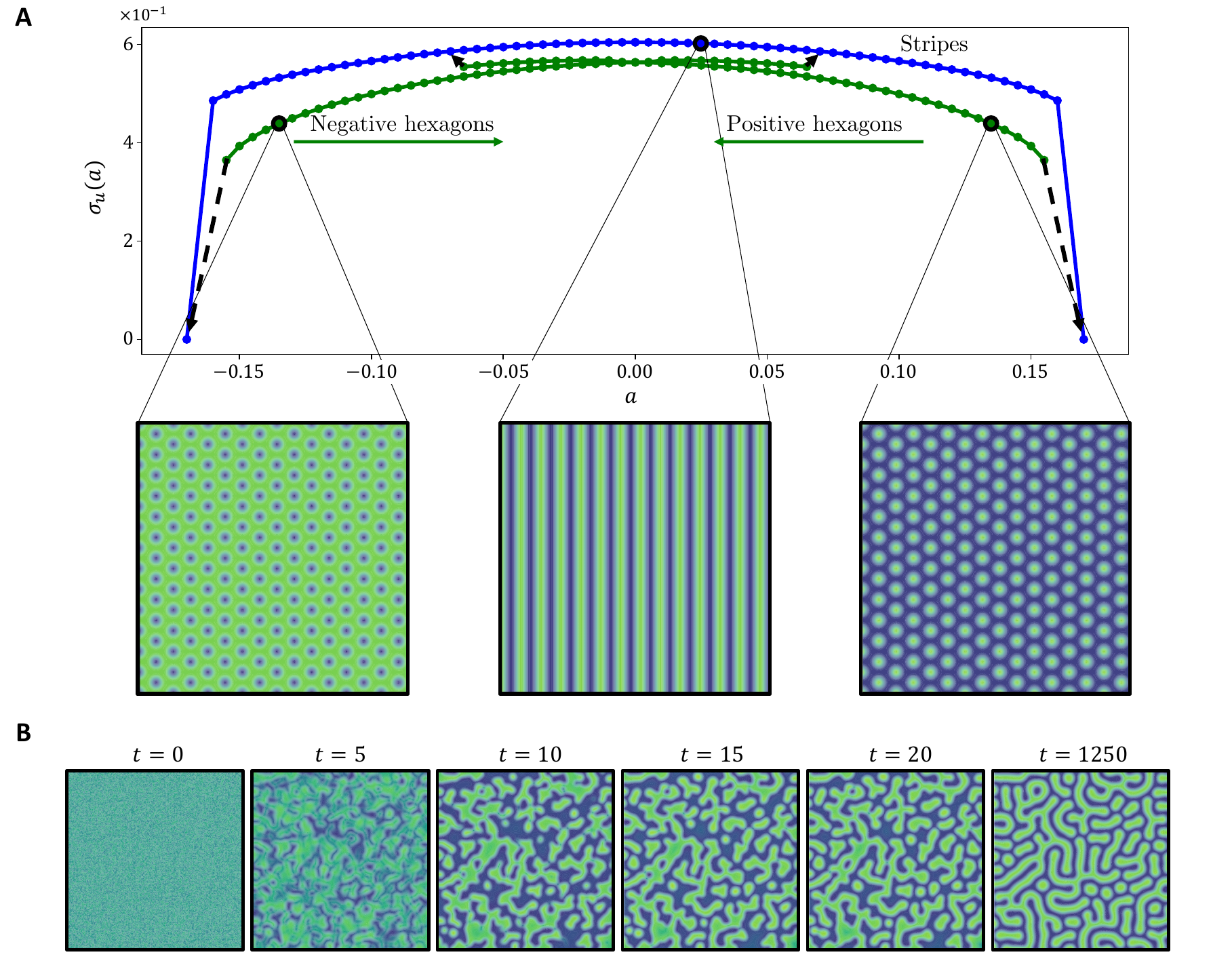}
\caption{\textbf{Creation of spatial steady patterns in the two dimensional FHN model.} \textbf{A}. Evolution of stripe and hexagonal patterns with respect to the control parameter $a$ for $(b,\varepsilon,D_u,D_v)$ equals to $(1.26,0.5,1,5)$. The patterns are initialized by perturbing the homogeneous system with a striped/hexagonal pattern signal with the critical wavelength of $(a,b)$ equals to $(0.025,1.26)$. All the domain exhibiting patterns also exhibits bistability of the stripe pattern with, either the positive hexagons $a\in(-0.070,0.165)$, or the negative hexagons $a\in(-0.165,0.070)$. Furthermore, the central region, $a\in(-0.07,0.07)$, exhibits bistability between the three patterns. \textbf{B}. Creation of a labyrinth pattern from random noisy initial conditions for the same initialization parameters used in the stripes, i.e.~$(a,b,\varepsilon,D_u,D_v)=(0.025,1.26,0.5,1,5)$.\label{fig:2D_Stationary_Patterns}}
\end{figure}

Mathematically, wave train studies transcend the FHN
context~\cite{Hagan1981,Dockery1988} due to their widespread applicability.
Utilizing the FHN model to describe wave trains, regularly used in cell cycle
contexts~\cite{75,nolet2020,ditalia2022,64,74,cebrian2024}, underscores its capacity to
mimic target patterns resembling the behavior of a nucleus or even multiple
nuclei within the cellular context~\cite{74,nolet2020,cebrian2024}. Such patterns and their response to time-scale separation and to the presence of spatial heterogeneity are not unique to the FHN model. They are also seen in chemical~\cite{Taylor2002}, cell cycle~\cite{chang2013mitotic,puls2024mitotic}, cardiac~\cite{Alonso_2016}, and neuronal models~\cite{Muller2018}, underscoring the FHN’s conceptual utility in exploring relaxation dynamics across various scientific domains, both theoretically and experimentally.
\begin{figure}
\includegraphics[width=\textwidth]{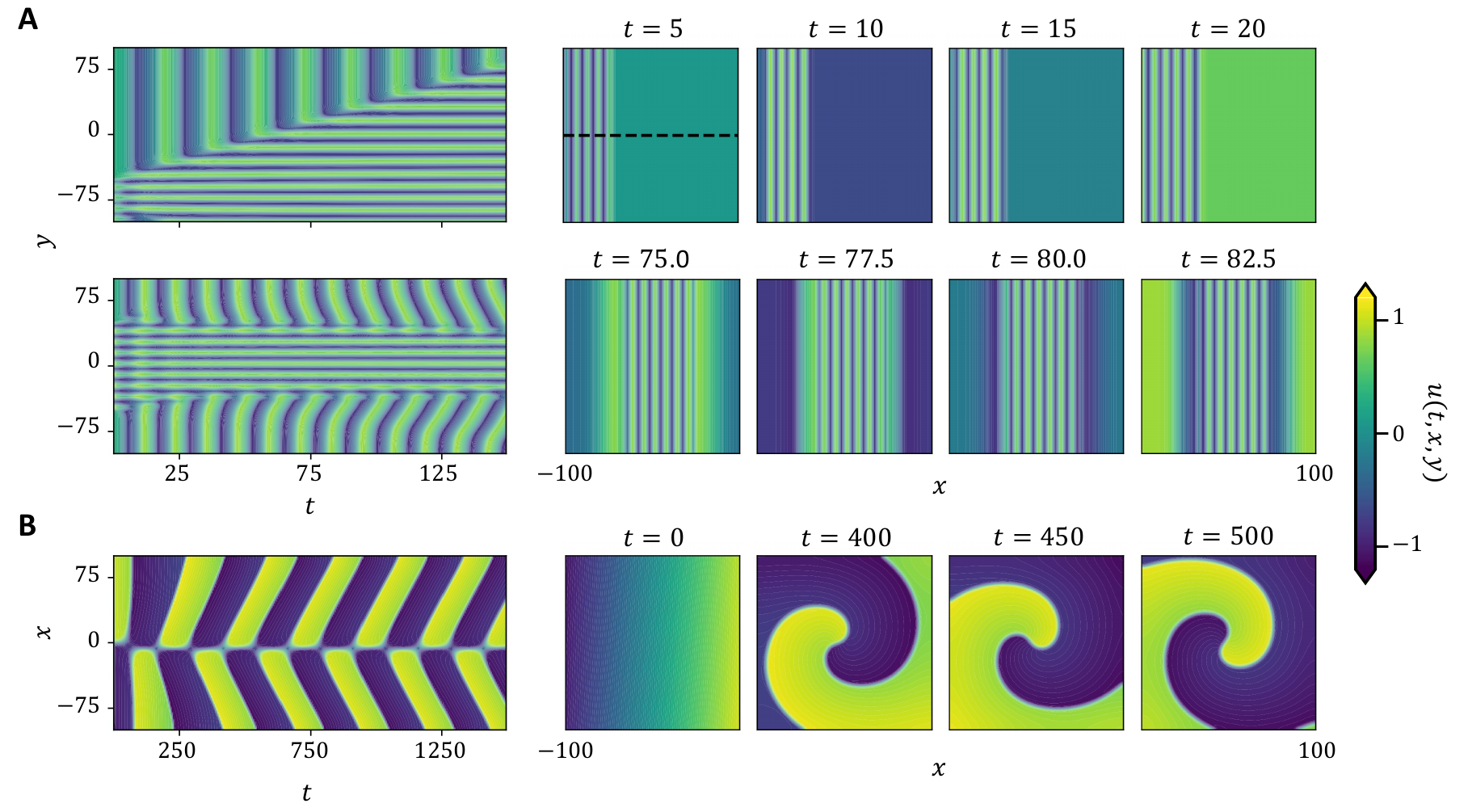}
\caption{\textbf{Dynamics of 2D spatio-temporal patterns.}
\textbf{A}. The dynamic interplay between the homogeneous state's oscillatory dynamics and spatial pattern is shown, set within the Turing-Hopf regime for parameters $(a,\varepsilon,D_u,D_v)=(0.025,0.5,1,5)$. In the upper series ($b=1.1$), a striped pattern emerges, generated cycle-by-cycle through periodic oscillations, which progressively takes over the oscillatory domain, leading to a fully established striped pattern across the system. Conversely, the lower series ($b=0.8$) shows a distinct dynamic equilibrium, where the system bifurcates into oscillatory and stationary patterned halves. At the interface, oscillatory dynamics verge on stabilization, while the adjacent striped pattern exhibits rhythmic fluctuations, akin to a `breathing stripe', highlighting a unique boundary behavior.
\textbf{B}. Spiral wave formation, originating from a phase defect induced by gradient-like initial conditions for $u$ (horizontal) and $v$ (vertical), with parameters $(a,b,\varepsilon,D_u,D_v)=(0.1,1,0.005,1,1)$ (oscillatory regime). A kymograph, representing a horizontal cross-section through the spiral's core, captures the phase defect. 
The numerical simulations in A and B use Neumann boundary conditions.\label{fig:2D_Temporal_Patterns}}
\end{figure}

\subsection{Complex spatio-temporal dynamics in two spatial dimensions}
\label{SS:pattern2D}

The coherent structures identified in the one-dimensional case serve as a
foundational classification for understanding the system's dynamics and the
mechanisms behind their formation. In higher spatial dimensions, the additional
degrees of freedom not only facilitate the emergence of more complex patterns
but also introduce new mechanisms such as curvature-driven
dynamics~\cite{hagberg1997,tyson1988,85,gomila2021curvature,avitabile2010} and
expanded parameter ranges for pattern existence, enhancing ecological
resilience~\cite{rietkerk2021}.

\subsubsection{Universal phenomena of pattern formation}

Pattern formation has been the subject of investigation in a wide variety of
contexts such as fluid dynamics, optics, morphogenesis, or
ecology~\cite{STspatial}. Despite the unique intricacies of each system, the
emergence of regular spatial structures is a universally observed phenomenon,
stemming from extended spatial interactions. This universality allows for a
unified mathematical framework to describe these
patterns~\cite{STspatial,walgraef2012spatio}. The general theory of pattern
formation suggests a predictable sequence of spatial organization changes in
response to variations in control parameters, typically progressing from
negative localized states to hexagons, stripes, and then positive localized
states~\cite{patterns_evolution}. \\

The FHN model, though fitting within this theoretical framework, has been less
extensively characterized in pattern formation compared to other
reaction-diffusion models, with most studies focusing on theoretical aspects
like amplitude equations, interface instabilities, or the effects of
cross-diffusion in population
dynamics~\cite{47,51,54,56,57,58,59,72,148,149,151,175}. This is possibly due
to its predominant association with excitability rather than spatial
self-organization.\\

The exploration of the bifurcation structure of localized states and patterns,
crucial for understanding a system's self-organizational response, has been
thorough in various spatially extended models but remains incomplete for the
FHN model. In biology, for instance, the morphogenetic patterns have been
examined using the Gierer-Meinhardt model~\cite{al2021unified}, while chemical
pattern formations have been delineated through the
Gray-Scott~\cite{al2021unified} and Brusselator~\cite{uecker2020snaking}
models. Ecological studies have extensively investigated vegetation patterns to
pinpoint tipping points critical for understanding phenomena like
desertification~\cite{patterns_evolution,gandhi2018spatially,zelnik2018implications,zelnik2013regime,meron2019multistability,al2023transitions,parra2020formation,ruizreynes2020,ruiz2020general}.

\subsubsection{Stationary patterns in two spatial dimensions}

In our exploration of the FHN model, we turn our attention to two spatial
dimensions to study the bifurcation diagram of spatial structures. While our
focus is on two dimensions, it is important to note that research has extended
into higher-dimensional pattern investigations, offering a richer understanding
of spatial dynamics~\cite{86,87}.\\

Stationary patterns, stemming from Turing bifurcations, have been a significant
area of study. Various researchers have looked into the genesis of these
spatial formations, formulating amplitude equations~\cite{51,72}. A notable
challenge is the close proximity of Turing and Hopf bifurcations, complicating
pattern identification. To circumvent this, strategies involve operating near
the Turing threshold or employing white noise to steer the system towards
stable pattern branches. Our numerical simulations reveal a diversity of stable
spatial patterns across different control parameters, specifically for
$b=1.26$, $\varepsilon=0.5$, $D_u=1$, and $D_v=5$, as illustrated in
Fig.~\ref{fig:2D_Stationary_Patterns}. These patterns, transitioning from negative
to positive spots through labyrinthine structures, underscore the labyrinth
configuration's robust stability for the chosen parameters. Goldstein's work on
interface growth within the FHN framework~\cite{57}, inspired by chemical front
interactions~\cite{lee1993pattern}, aligns with our findings in the labyrinth
domains (Fig.~\ref{fig:2D_Stationary_Patterns}B).

\subsubsection{Pattern transitions and secondary instabilities}

An intriguing aspect of pattern formation is the transition between various
structures, often marked by coexistence regimes where different patterns vie
for dominance, influenced by their relative stability. This dynamical
interplay, further complicated by secondary instabilities like Eckhaus or
ZigZag distortions~\cite{72}, is depicted through the mean difference relative
to the homogeneous state ($u_0$), highlighting a stark contrast between
compact structures like spots and expansive ones like labyrinths. Our
simulations also indicate a hysteresis phenomenon during transitions from
hexagonal to labyrinthine patterns, marked by dashed black arrows in
Fig.~\ref{fig:2D_Stationary_Patterns}A. This behavior, aligning with Kuznetsov's
findings~\cite{89}, confirms that the observed patterns adhere to the
theoretical framework of pattern
formation~\cite{patterns_evolution,walgraef2012spatio}, sharing commonalities
with reaction-diffusion models across various scientific fields.

\subsubsection{The Turing-{Hopf} regime}

Beyond stationary patterns, spatio-temporal structures in the FHN model offer a
rich set of dynamic behaviors. These structures can manifest as either
transient phenomena or stable configurations, distinguished by their origins
from the Turing-Hopf interaction~\cite{72} or purely Hopf instabilities.\\

The interplay between oscillatory dynamics and pattern formation in the
Turing-Hopf regime introduces a diverse array of spatio-temporal patterns,
including expanding rings, pulsating waves, oscillatory patterns, and
self-replicating spots~\cite{56,59,86}. 
This regime, primarily conceptualized through the Ginzburg-Landau
equation~\cite{walgraef2012spatio}, remains less explored in the FHN model,
presenting an opportunity to deepen our understanding of systems where pattern
formation and oscillatory behaviors
coexist~\cite{56,brauns2021bulk,lee1994experimental,175,hladyshau2021spatiotemporal}. 
Phenomena such as 'breathing stripes' exemplify the complex dynamics within this regime, as illustrated in  Fig.~\ref{fig:2D_Temporal_Patterns}A. Here, we show two scenarios: one where an initial pattern progressively dominates the oscillatory landscape and another featuring a dynamic equilibrium between oscillatory and patterned regions, highlighted by the breathing stripe at the boundary.

\subsubsection{Spiral waves}

Spiral waves have been widely studied in the FHN
model for mathematical
purposes~\cite{38,43,44,47,54,68,69,76,85,86,88,109,129,171} and for their
applicability in cardiology~\cite{66,67,80,93,102,124,130}, where spiral
behavior has been associated with arrhythmias. From the simple spirals studied
by Erhaud and Winfree~\cite{54,76} to the complex collisions studied by the
latter~\cite{69}, and through spiral turbulence emerging from the loss of
stability of these structures~\cite{87,88}, the FHN model has been utilized to
characterize several regimes associated with spiral behavior. In
Fig.~\ref{fig:2D_Temporal_Patterns}B, we illustrate the creation of a stable spiral
arising from a phase defect induced by considering horizontal gradient-like
initial conditions for $u$ and horizontal gradient-like initial
conditions for $v$ in an oscillatory regime ($a=0.1$,
$b=1$, $\varepsilon=0.005$, $D_u=1$, $D_v=1$).\\

Spiral wave patterns, particularly relevant in cardiological contexts due to
their association with arrhythmic behaviors~\cite{66,67,80,93,102,124,130},
represent another facet of the FHN model's versatility. From the foundational
works on simple spirals~\cite{54,76} to the studies on spiral
turbulence~\cite{87,88} and collision dynamics~\cite{69}, the model serves as a
vital tool in understanding the nuances of spiral behavior.
Fig.~\ref{fig:2D_Temporal_Patterns}B demonstrates the formation of a stable spiral,
a phenomenon induced by specific initial conditions (here a phase defect)
within an oscillatory setting. This insight into spiral formation within the
FHN model not only advances mathematical
explorations~\cite{38,43,44,47,54,68,69,76,85,86,88,109,129,171} but also has
profound implications in the study of cardiac
arrhythmias~\cite{66,67,80,93,102,124,130}, offering potential avenues for
therapeutic interventions.

\section{Discretely coupled {FitzHugh}-Nagumo equations}
\label{discrete}

\begin{figure}
\includegraphics[width=\textwidth]{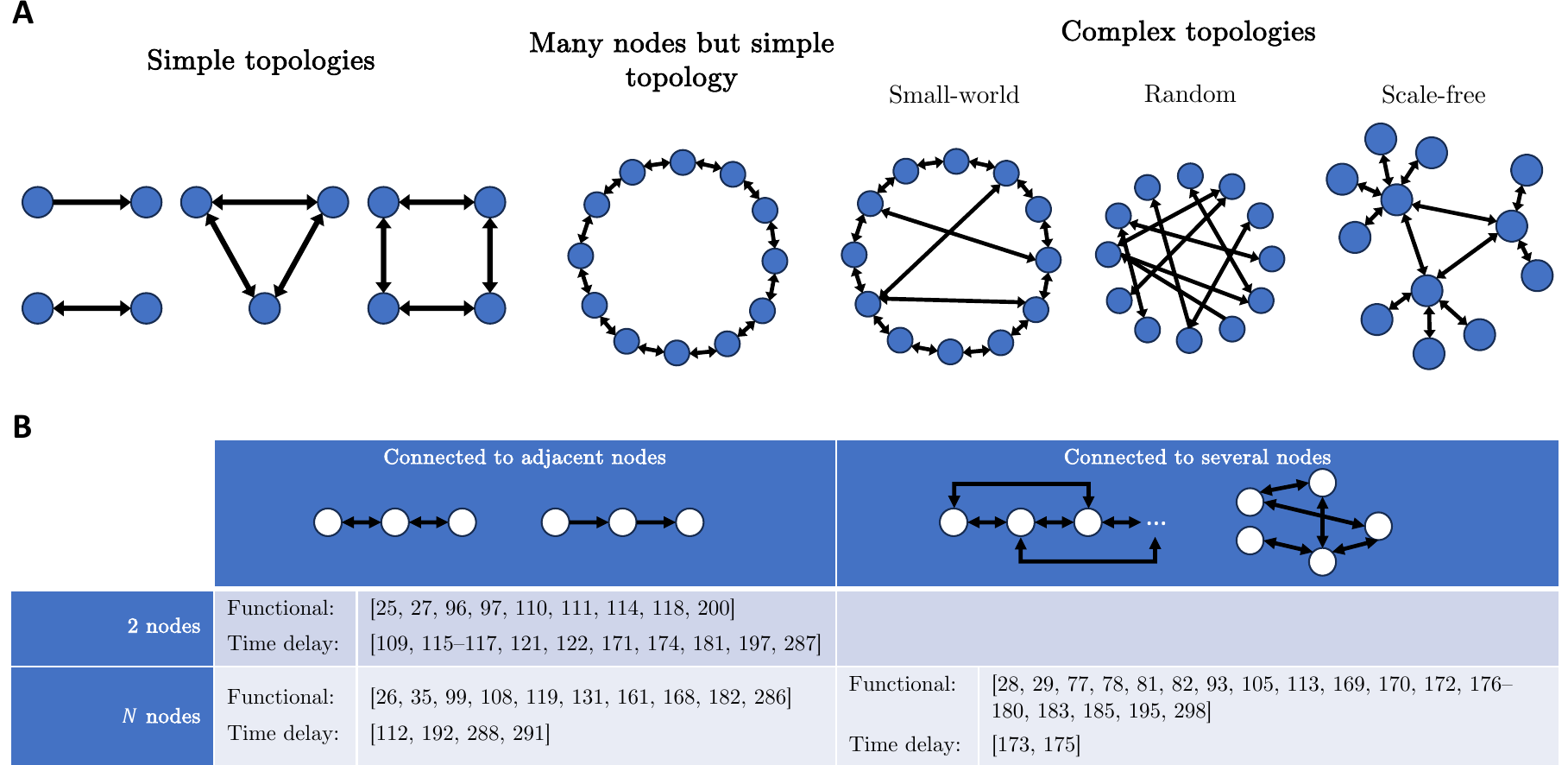}
\caption{\textbf{Different network topologies and coupling functions in discretely coupled FHN equations.}
The complexity and variety of dynamics in coupled ODEs are influenced by the coupling mechanisms and the network's structure. This figure provides an overview of common network topologies and categorizes existing FHN literature based on network topology and coupling types.
\textbf{A}. Network topologies, ranging from simple unidirectional links between two nodes to complex small-world networks characterized by a few highly connected nodes amid many with fewer connections. The diversity of topologies is vast, and these examples barely scratch the surface, especially when considering the myriad coupling functions that interconnect these nodes.
\textbf{B}. Existing research on discretely-coupled FHN equations is delineated based on the specific network architectures and the nature of the coupling employed.\label{fig:topologies}}
\end{figure}

Discretely coupled ODEs, describing various network configurations, have
garnered substantial attention in diverse fields such as
neuroscience~\cite{7,96,128,140,182}, electrical
systems~\cite{112,138,161,169,176,119}, and biology~\cite{28,145,146,178},
owing to their relevance and the broad spectrum of topologies and coupling
terms they accommodate. This versatility allows for the exploration of various
phenomena, enriching the mathematical study of
networks~\cite{9,10,12,19,23,24,38,92,104,107,108,118,120,121,122,123,126,131,141,165,172,182,185}.\\

Network topology, concerning the structure and connectivity of nodes (Fig.~\ref{fig:topologies}A), combined with the diversity in coupling terms -- from simple functional forms to time-delayed terms -- sets the stage for endless modeling possibilities. The classification of networks, as per the literature, often revolves around the network size, the extent of node connections, and the nature of coupling terms, encompassing both functional forms and time delays (Fig.~\ref{fig:topologies}B).\\

The coupled FHN models have primarily been employed to investigate
synchronization
phenomena~\cite{7,9,24,28,38,92,96,104,107,108,118,121,122,123,126,128,131,138,140,141,161,165,169,178,182,185,8,26,27,111,114,115,116,117,119,125,162,179},
including the study of chimera
states~\cite{107,108,114,115,118,121,122,140,165}. Additionally, they have been
used to explore stability
properties~\cite{10,19,23,112,120,138,145,146,161,185,8,11,20,21,22,162,179},
traveling waves~\cite{71,145,146,16,115}, pattern
formation~\cite{38,92,172,12,176,162,104,120,123,165,115}, and the influence of
noise on dynamics~\cite{24,27,92,104,111,118,123,126,141,172}.\\

Given the vast array of possibilities with coupled ODEs, this review will focus on selected popular topics within the FHN model context. 
In  \ref{CoupledODE:subsection2}, we discuss the transition from continuous (PDE) systems to discrete coupled ODEs.
 \ref{CoupledODE:subsection3} looks into one of the most basic forms of
coupling -- the functional difference -- and focuses on a two-node
system~\cite{10,23}. This part provides insights into stability analysis and
synchronization dynamics, employing the simplest coupling scenarios frequently
found in literature~\cite{9,10,19,23,24,38,71,92,104,112,126,128,185}.
Finally,  \ref{CoupledODE:subsection4} discusses so-called chimera states, a key
area of interest within the FHN model studies, by replicating and discussing
the findings of Omelchenko \textit{et al.}~\cite{108,122}. This section underscores the
significance of chimeras in the study of complex network dynamics.

\subsection{From continuously to discretely coupled {{FHN}} systems}
\label{CoupledODE:subsection2}

The FHN model, initially conceived as a simple ODE~\cite{1}, was quickly
expanded into a PDE version to model pulse propagation in axons~\cite{2}. Here,
we characterize how discretely coupled FHN systems can approximate the behavior
of their continuous counterparts.\\

In computational simulations, the inherently finite nature of the number of
units is similar to the approach taken by Nagumo's electronic implementation. A
standard method for simulating spatial dynamics involves approximating the
Laplacian operator using finite difference schemes~\cite{FDIrregular},
expressed as:
\begin{equation}
u_{xx}\approx\dfrac{-2u(x)+u(x+\Delta x)+u(x-\Delta x)}{\Delta x^2},
\end{equation}
where $\Delta x$ represents the discretization step size.\\

Alternatively, networks might adopt various topologies based on functional
differences, influencing the coupling
dynamics~\cite{9,10,19,23,24,38,71,92,104,112,126,128,185}. A common form of
expressing this coupling is:
\begin{equation}\begin{array}{ @{} l  @{} }  {u_i}_t=f(u_i,v_i)+\eta(-2u_i+u_{i+1}+u_{i-1}), \vspace{3pt}\\
        {v_i}_t=\varepsilon g(u_i,v_i)+\eta(-2v_i+v_{i+1}+v_{i-1}),
\end{array}
\end{equation}
highlighting the functional dependency between a node and its neighbors. This discrete coupling closely resembles the discretized diffusion operator, with the diffusion coefficient correlating to both the coupling strength ($\eta$) and the discretization step.\\

\begin{figure}
\includegraphics[width=\textwidth]{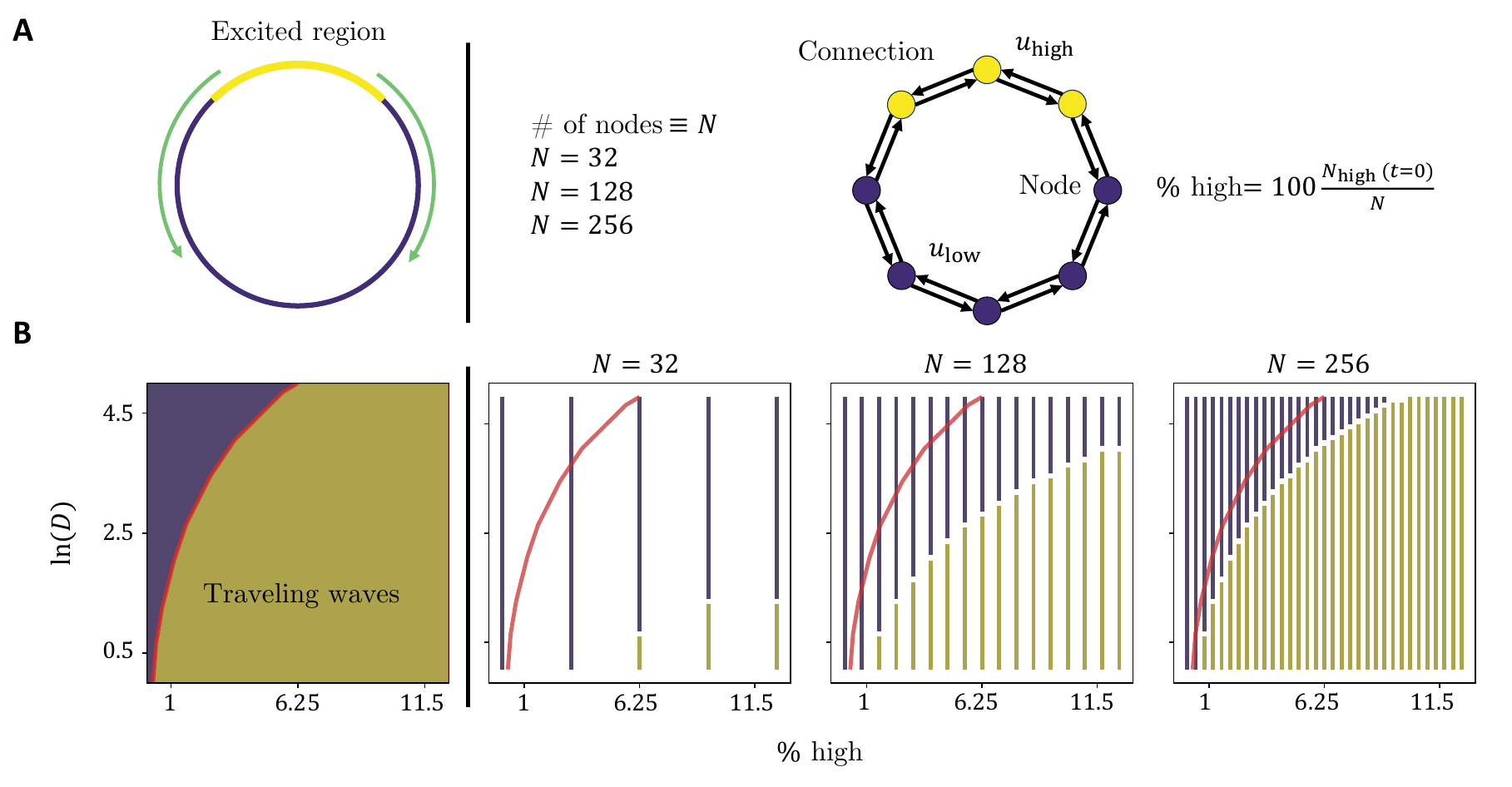}
\caption{\textbf{Transition from continuous to discrete systems: quantifying wave propagation.}
    \textbf{A.} A continuous system with periodic boundary conditions (left) is contrasted against its discrete counterpart (right). The presence or absence of traveling waves depends on whether the excited region propagates.
    \textbf{B.} Quantifying wave propagation when transitioning from a continuous system (left) to discrete systems (right), under parameters set to favor the active state ($a=-0.1$, $b=2$, $\varepsilon=0.01$). As the number of coupled nodes increases (32, 128, and 256 nodes examined), the region supporting wave propagation expands, progressively mirroring the continuous system's behavior. The activation percentages are delineated by the proportion of initially active nodes.\label{fig:transition}}
\end{figure}

An intriguing aspect is how wave propagation changes with the number of FHN
units or discretization step. Keener's work on simulating traveling waves in a
cardiac model using discrete FHN nodes serves as a notable example~\cite{71},
alongside other diverse applications~\cite{71,145,146,16,115}.
Fig.~\ref{fig:transition} shows the division in a continuous system (with a high
node count) between regions exhibiting traveling waves and those that do not,
based on the diffusion coefficient $D$ and the size of the initially
excited region (\% high). This numerical experiment is repeated for discrete
node counts of $N=(32,128,256)$. The results clarify that as the number of FHN
nodes increases, the threshold for wave propagation converges to that of the
continuous system, illustrating the transition between discrete and continuous
regimes.

\subsection{A unidirectional functional difference between two coupled {{FHN}} models}
\label{CoupledODE:subsection3}

A significant portion of the literature employing functional coupling consists
of functional differences, i.e.,~coupled ODEs where the differential equation
for the $i$th node is proportional to $\sum_j(u_i-u_j)$. Here, we explore
the dynamics of a simple two-node system through general functional forms and
examine the specific scenario of unidirectional coupling, similar to the
studies by Hoff \textit{et al.}~\cite{10} and Campbell \textit{et al.}~\cite{23}.
Campbell \textit{et al.}  assumed no coupling from node 2 to node 1.

\begin{figure}
\includegraphics[width=\textwidth]{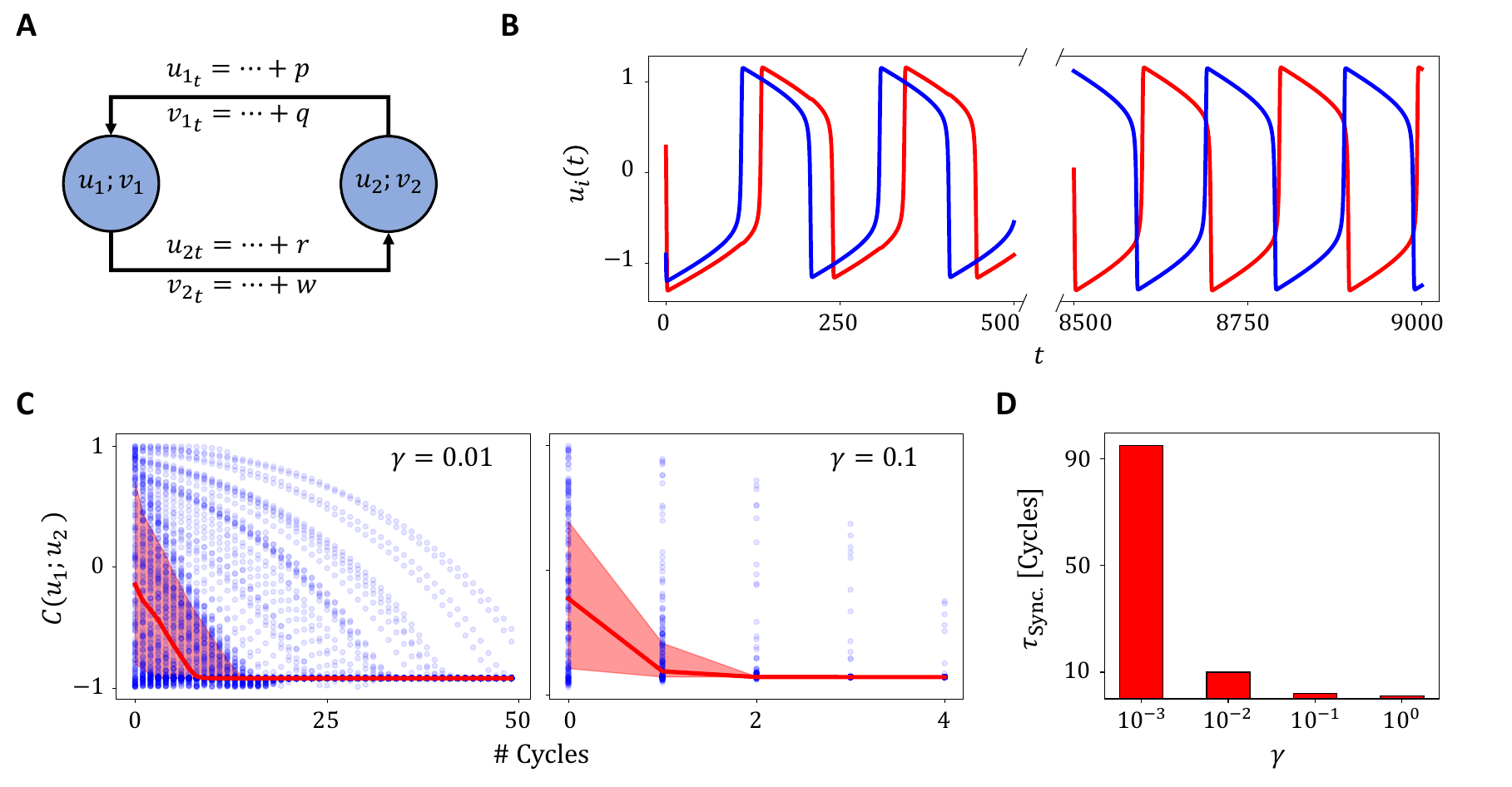}
\caption{\textbf{Dynamics of two coupled FHN nodes} \textbf{A}. The basic setup for two coupled FHN nodes. \textbf{B}. Time series of two interconnected FHN nodes operating in an oscillatory regime with parameters $(a,b,\varepsilon)=(0,0.5,0.01)$. Weak coupling is defined by $f=\gamma(u_1-u_2)$, with $g=h=m=0$ and $(\gamma=0.01)$. \textbf{C}. Correlation $C(u_1;u_2)$ in oscillatory cycles between the nodes over 200 iterations, starting from random initial conditions. \textbf{D}. The number of cycles required for the phase synchronization to stabilize at a median value, highlighting the dependency of this convergence on the coupling intensity.\label{fig:Time_Coupled}}
\end{figure}

\subsubsection{Stationary solutions of two coupled nodes}

For a two-node system (Fig.~\ref{fig:Time_Coupled}A), the stability equations are
derived considering general functions $p$, $q$, $r$,
and $w$. Focusing on functional differences, we model the scenario
where node 2 influences node 1 without reciprocal interaction, embodying the
`slave and master' dynamic prevalent in neuroscience~\cite{24}. The system
equations are given as:
\begin{equation}\begin{array}{ @{} l  @{} }        {u_1}_t=-u_1^3+u_1-v_1+p(u_1,v_1,u_2,v_2),\vspace{3pt}\\
        {v_1}_t=\varepsilon(u_1-bv_1+a)+q(u_1,v_1,u_2,v_2),\vspace{3pt}\\
        {u_2}_t=-u_2^3+u_2-v_2+r(u_1,v_1,u_2,v_2),\vspace{3pt}\\
        {v_2}_t=\varepsilon(u_2-bv_2+a)+w(u_1,v_1,u_2,v_2),
\end{array}
    \label{EqCoupledEx}
\end{equation}
with $p(u_1,v_1,u_2,v_2)\equiv\gamma(u_1-u_2)$, $q(u_1,v_1,u_2,v_2)=r(u_1,v_1,u_2,v_2)= w(u_1,v_1,u_2,v_2)=0$.\\

Analyzing the nullclines and fixed points reveals the potential for complex dynamics, extending to multiple stable states under simple unidirectional coupling.
Such analysis hinges on the specific functional forms involved, often leading to equations for which there are no analytical solutions. Thus, we propose general solutions of the form $(u_{\text{FP},1},v_{\text{FP},1},u_{\text{FP},2},v_{\text{FP},2})$. In our case, although the equations are not analytically solvable, we derive certain simplified relational expressions:
\begin{equation}\begin{array}{ll}
    v_{\text{FP},1}=\frac{u_{\text{FP},1}+a}{b}, & bu_{\text{FP},1}^3 + (1-b-b\gamma)u_{\text{FP},1} + b\gamma u_{\text{FP},2},\vspace{3pt}\\
    v_{\text{FP},2}=\frac{u_{\text{FP},2}+a}{b}, & bu_{\text{FP},2}^3 + (1-b)u_{\text{FP},2} + a = 0,
\end{array}
\end{equation}
where the second equation represents node 2 as operating independently, similar to the scenario without coupling. Conversely, the dynamics of node 1 now exhibit a dependency on $u_2$, which could result in mono- or tri-stable behaviors, potentially leading to as many as nine stable fixed points. This elementary derivation highlights how the model's complexity increases as the power of $3^N$, where $N$ is the number of nodes, assuming a simplistic unidirectional coupling scheme and gauging complexity by the number of fixed points.

\subsubsection{Linear stability analysis}
Continuing with this analytical approach, we now explore the stability of fixed points by considering small perturbations in the form of  
\[
(u_1, v_1, u_2, v_2) = (u_{\text{FP},1}, v_{\text{FP},1}, u_{\text{FP},2}, v_{\text{FP},2}) + \epsilon (\xi_{u_1}, \xi_{v_1}, \xi_{u_2}, \xi_{v_2}) e^{\sigma t} + c.c.,
\]  
where $\epsilon$ is sufficiently small. This leads to an eigenvalue problem described by:  
\begin{equation}
    (J - \sigma I_{4 \times 4})
    \begin{pmatrix}
        \xi_{u_1} \\
        \xi_{v_1} \\
        \xi_{u_2} \\
        \xi_{v_2}
    \end{pmatrix}
    =
    \begin{pmatrix}
        0 \\
        0 \\
        0 \\
        0
    \end{pmatrix},
\end{equation}  
where the Jacobian $J(u_1, v_1, u_2, v_2)$ is given by:  
\[
    J \equiv
    \begin{pmatrix}
        -3u_{\text{FP},1}^2 + 1 + \gamma & -1 & -\gamma & 0 \\
        \varepsilon & -\varepsilon b & 0 & 0 \\
        0 & 0 & -3u_{\text{FP},2}^2 + 1 & -1 \\
        0 & 0 & \varepsilon & -\varepsilon b
    \end{pmatrix}.
\]

Unlike two-dimensional systems, the eigenvalues cannot be calculated solely using the trace and determinant. Instead, computational or advanced analytical methods are required to solve the characteristic polynomial:  
\begin{equation}
\text{det}(J - \sigma I_{4 \times 4}) = \left[\sigma^2-(\kappa_1-\varepsilon b)\sigma+\varepsilon-\kappa_1\varepsilon b\right]\left[\sigma^2-(\kappa_2-\varepsilon b)\sigma+\varepsilon-\kappa_2\varepsilon b\right],
\end{equation}
where $\kappa_1=-3u_{\text{FP},1}^2 + 1 + \gamma$ and $\kappa_2=-3u_{\text{FP},2}^2 + 1$. Analogous to the zero-trace condition, a Hopf bifurcation occurs when a pair of eigenvalues becomes purely imaginary, while the remaining eigenvalues have negative real parts. Thus, $\kappa_i-\varepsilon b=0$ and $1-\kappa_ib>0$
for $i=1,2$, while $\mathbb{R}[\kappa_j-\varepsilon b\pm\sqrt{(\kappa_j-\varepsilon b)^2-4\varepsilon(1-\kappa_jb)}]\leq0$ for $j\neq i$.

This derivation can be extended to other coupling types, as illustrated in Fig.~\ref{fig:topologies}B. In certain scenarios, the symmetric properties of the coupling and the fixed points can simplify the analysis, yielding more straightforward analytical expressions~\cite{23}. For those interested in numerical bifurcation analysis in ODE systems, we recommend the work of Dhooge \textit{et al.}~\cite{BifurcationNumerical}, which offers a comprehensive bifurcation analysis toolkit~\cite{10}.

\subsubsection{Visualizing the coupling effects}

In  Fig.~\ref{fig:Time_Coupled}B, the impact of coupling on oscillatory dynamics
between two FHN nodes is shown, highlighting a tendency towards antiphase
synchronization. To quantify this synchronization, we determine the \textit{characteristic synchronization time}
for different synchronization strengths (Fig.~\ref{fig:Time_Coupled}C).
This approach involves analyzing the correlation across individual oscillations
and tracking the evolution of this correlation over successive cycles, aligning
with techniques described by Toral \textit{et al.}~\cite{24}. This reveals
that the number of cycles required to reach the ultimate antiphase
synchronized state
varies with the coupling strength (Fig.~\ref{fig:Time_Coupled}D).\\

Despite the model's simplicity, it encapsulates a broad spectrum of
dynamics~\cite{9,10,19,23,24,38,71,92,104,112,126,128,185}, setting the stage
for exploring more complex systems involving multiple oscillators and
connections, often used in the study of chimera states. 

\begin{figure}
\includegraphics[width=\textwidth]{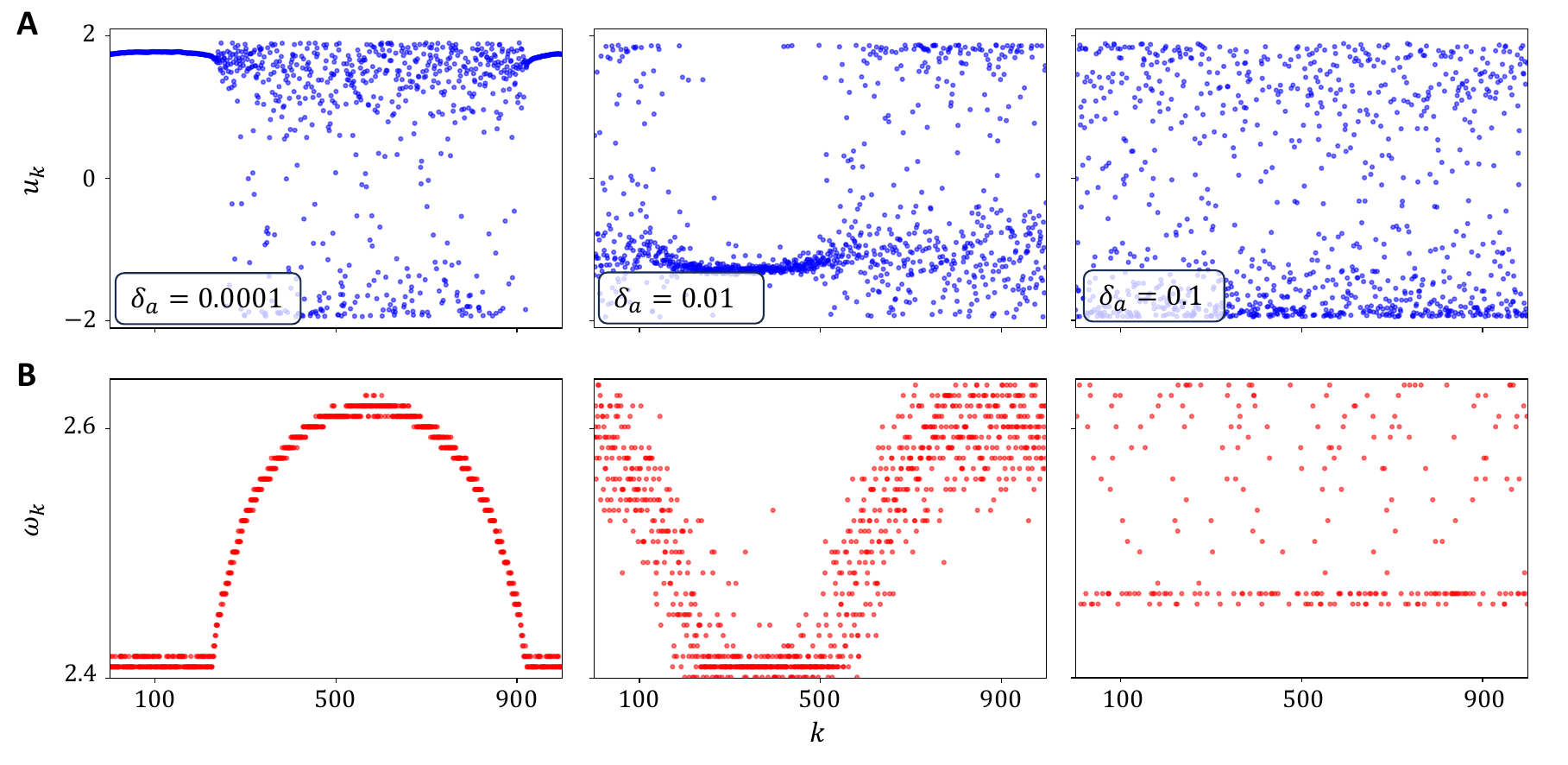}
\caption{\textbf{Chimera states in discretely coupled FHN oscillators.} Parameters set for the simulation include $a_m=0.5$, coupling strength $\sigma=0.1$, time scale separation $\varepsilon=0.05$, coupling phase $\phi=\pi/2-0.1$, with a network size of $N=1000$ and coupling range $R=350$.  \textbf{A} Spatial distribution of the oscillator variable $u_k$, showing regions of synchronization contrasted with areas of asynchronization. The impact of increased parameter variability ($\delta_a$) progressively blurs the distinction between these regions. \textbf{B}.  Mean phase velocities ($\omega_k$) further characterize the asynchronization, illustrating the dynamical complexity and the sensitivity to initial conditions and parameter heterogeneity inherent in chimera states.\label{fig:chimera}}
\end{figure}

\subsection{Chimera states}
\label{CoupledODE:subsection4}

Chimera states, known for their intriguing spatio-temporal patterns characterized by coexisting domains of coherence and incoherence, have garnered significant attention in studies of coupled 
FHN oscillators~\cite{107,108,114,115,118,121,122,140,165,huang2024frequency}.
This dynamical regime can manifest across diverse network configurations. Work
by Omelchenko \textit{et al.}~\cite{108,122} offers valuable insights into the underlying
mechanisms and conditions conducive to chimera states.\\

Contrary to the assumption that complex networks are essential for chimera states, Omelchenko \textit{et al.}  demonstrate their emergence within a relatively simple setup: a ring network of $N$ FHN units with symmetric connections but non-uniform parameters. The dynamics of each node in the network are described by:
\begin{equation}\begin{array}{ll}
     \varepsilon{u_k}_t=-\frac{u_k^3}{3}+u_k-v_k+\frac{\sigma}{2R}\sum_{k-R}^{k+R}[b_{uu}(u_j-u_k)+b_{uv}(v_j-v_k)], \vspace{3pt}\\ 
  {v_k}_t=u_k+a_k+\frac{\sigma}{2R}\sum_{k-R}^{k+R}[b_{vu}(u_j-u_k)+b_{vv}(v_j-v_k)],  
\end{array}
\end{equation}
with $R$ representing the radius of interaction among adjacent nodes.
Following~\cite{108,122} we chose $b=0$, $\varepsilon=0.05$ and $\sigma=0.1$.
Furthermore, the coupling terms of the equation are defined by a
rotation matrix,
\begin{equation}
\begin{pmatrix}
        b_{uu} & b_{uv} \\
        b_{vu} & b_{vv}
\end{pmatrix}
=
\begin{pmatrix}
        \cos{\phi} & \sin{\phi} \\
        -\sin{\phi} & \cos{\phi}
\end{pmatrix}
,
\end{equation}
with cross couplings $b_{uv}$ and $b_{vu}$. \\

Chimera states are identified under specific initial conditions and parameter
values. Nodes are initialized with random values constrained on a circle of
radius $2$ in the phase space, ensuring diverse starting points.
Heterogeneity is introduced through the parameter $a_k$, assigned values
from a Gaussian distribution centered around $a_m=0.5$ with a standard
deviation $\delta_a$, determining the presence of chimera states. Moreover,
the angle considered for the coupling terms must be close to $\pi/2$, as
off-diagonal terms need to be enhanced~\cite{108}. 
A parameter diagram for the set $(\sigma,R)$ for which one can encounter
chimeras is provided in~\cite{108}.\\

 Fig.~\ref{fig:chimera} illustrates the manifestation of chimera states. The
analysis often involves mean phase velocities, indicative of the system's
dynamical state. In these states, a segment of incoherently oscillating nodes
coexists with a synchronized group, forming distinct coherent and incoherent
domains in the network. This phenomenon aligns with findings from Omelchenko
\textit{et al.}~\cite{122}, where increasing initial variation gradually merges the two
distinct regions.\\

Alternative configurations, such as networks with non-uniform coupling, can
also lead to chimera states. This aspect is particularly relevant to
neuroscience, where inhomogeneous connectivity is observed in the mammalian
brain, suggesting a potential link between chimera states and brain
dynamics~\cite{Vesna2014}. The exploration of chimera states continues to be a
rich field of study, with diverse methodologies and applications expanding our
understanding of complex systems.\\

In summary, the study of coupled oscillators unveils a vast range of dynamic
behaviors, attributed to the diverse network structures, coupling mechanisms,
and the inherent parameter versatility within these systems. For simpler configurations, analytical insights into the system's
bifurcation and stability aspects can be obtained~\cite{10,23}. The application
of coupled oscillator models extends across various fields, in modeling
neuronal~\cite{7,96,128,140,182}, electrical~\cite{112,138,161,169,176,119},
and biological systems~\cite{28,145,146,178,cebrian2024}. A significant portion of research
focuses on understanding synchronization patterns across
nodes~\cite{107,108,114,115,118,121,122,140,165}, often employing correlation
functions for analysis~\cite{24}. Additionally, the role of noise in
influencing system dynamics receives considerable attention, revealing its
impact on the coupled oscillators'
behavior~\cite{24,27,92,104,111,118,123,126,141,172}.

\section{Conclusions}

In this work, we explored the most prominent dynamical behaviors in the FitzHugh-Nagumo (FHN) model, a framework initially developed in neuroscience that has since broadened its application across various scientific fields. Beyond its origins, the FHN model has become instrumental in elucidating phenomena spanning from cardiac dynamics to mathematical and physical concepts, underscoring its adaptability and significance (Figs.~\ref{fig:PreviosReferences} and \ref{fig:HHmodel}D).\\

Within neuroscience, the FHN model has been crucial for understanding neuronal dynamics and the interplay within neural networks, offering insights into synchronization, coordinated activity patterns, and the propagation of action potentials along axons. Similarly, in cardiology, the FHN model serves as a powerful tool for simulating cardiac behavior, exploring phenomena like spirals, and investigating conditions such as ventricular fibrillation. Its utility spans other biological phenomena too, such as cell cycle dynamics and cytosolic calcium fluctuations.  Moreover, outside of biology, the FHN model is applied in electronic circuit designs mimicking neuronal activity, innovations in all-optical spiking neurons, and contributes to the broader scientific understanding in mathematics and physics, especially within excitable systems.\\

We structured our analysis into three primary sections. Firstly, we examined the original FHN model [Eq.~\ref{EqTemporal}], discussing widely observed dynamical regimes such as monostability, multistability, relaxation oscillations, and excitability. We examined the role of local and global bifurcations in shaping these regimes, emphasizing the importance of time scale separation.
Secondly, we explored the diffusively coupled FHN model [Eq.~\ref{EqSpatial}], introducing spatial coupling through diffusion. Through theoretical analysis, we investigated stationary homogeneous solutions, their linear stability, and spatially structured dynamic solutions, including traveling structures and spatially extended patterns. We studied the emergence of a Turing instability and the resulting spatially structured Turing patterns. Additionally, we examined front solutions, localized states, traveling pulses, and pacemaker-driven waves within the oscillatory domain, highlighting the richness of patterns that arose in different spatial dimensions.
Lastly, we explored discretely coupled FHN equations [Eq.~\ref{EqCoupled}]. This is the broadest category as here one can consider a multitude of different network topologies and coupling terms. We focussed on synchronization properties in two coupled FHN modules, the existence of traveling waves when transitioning from continuous diffusive coupling to discrete coupling, and the emergence of chimera states characterized by spatio-temporal patterns of coherent and incoherent behavior.\\

In conclusion, we hope our review will serve as a guide for understanding and using the diverse dynamical behaviors offered by the FHN model. Throughout our analysis, stability analyses and bifurcation studies provided insights into the observed dynamics. By exploring its applications across multiple disciplines, we aimed to inspire further exploration and application of the FHN model in diverse scientific domains.

\section{Discussion}

The dynamics of the FHN model are influenced by various factors, including external driving, internal feedback, inter-system coupling, noise, and time delays. These elements contribute not only to the model's inherent complexity but also to its extensive applicability across physical and biological domains. While our review has examined the dynamics of single and coupled FHN systems in some detail, it has not extensively covered the roles of external driving, self-feedback, time delays, and noise. Each of these aspects warrants further exploration for a comprehensive understanding of the FHN model's dynamics and potential.
Living systems are typically non-equilibrium, active, and multi-scale complex systems that often operate in transient states away from stable attractors. The FHN model is an excellent candidate to serve as a testbed for developing new methods to analyze such transient dynamics and non-reciprocal interactions. Integrating the FHN model within multiphysics frameworks, which encapsulate diverse physical interactions, is essential for advancing our understanding of these systems. Such new techniques and integrative models are particularly important in the complex simulation of biological phenomena. Cardiac and neuronal examples, in particular, highlight the FHN model's adaptability and significance in these areas.\\

\textbf{External Driving.} Introducing an external stimulus to the FHN system
can induce resonance or forced oscillations, when the stimulus frequency aligns
with the system's inherent frequency. This aspect is pivotal in understanding
biological rhythms and has applications in electronic and mechanical systems,
as well as in neuroscience for modeling brain rhythms and treating disorders
like Parkinson's. Researchers have looked into the dynamics of externally
driven FHN models, revealing their potential for controlling excitable
behaviors~\cite{106,155,167}. Of course, the external driving could also come
from another FHN system, leading to two autonomous dynamical systems in a
master-slave configuration. Synchronization has been studied in such
unidirectionally coupled models, also in the presence of additional external
aperiodic forcing~\cite{119}.\\

\textbf{Feedback.} Feedback, where the system's output loops back as an input,
can have a large effect on the system dynamics. For example, oscillations can
be stabilized or destabilized. Positive feedback amplifies oscillations, useful
in simulating biological rhythms like heartbeats, while negative feedback
maintains stability, crucial for homeostasis in biological systems. The FHN
model's incorporation of feedback offers a detailed simulation of biological
excitability, with implications for understanding and treating oscillatory
disruptions in neurological and cardiac conditions. Time-delayed feedback, in
particular, adds complexity to the system's dynamics, influencing neuron
synchronization and offering pathways to innovative optical pulse generation
and neuromorphic computing applications~\cite{110,113,125}.\\

\textbf{Time delays.} Time delays play an important role in nonlinear systems, extensively researched due to their pervasive presence across a spectrum of physical and biological contexts. These delays, which can be intentionally introduced via feedback loops or inherently present in various systems, play a significant role in shaping the behavior of dynamical models. In systems ranging from optical and electronic devices to neuronal networks and technological infrastructures, time delays emerge due to the finite speed of signal transmission, processing times, and the effects of memory and latency. 
Incorporating time delays into models offers new strategies for controlling and
designing nonlinear systems, enhancing their stability and performance. In
neural networks, for instance, the inherent time delays in signal transmission,
owing to diverse neural pathways, introduce complex dynamics that can be
replicated in models like the FHN framework, not only through delayed
interactions but as an intrinsic characteristic of the system. These delays can
significantly impact the network's behavior, leading to phenomena such as
enhanced synchronization, the suppression of undesirable oscillations, and the
formation of chimera
states~\cite{117,162,179,21,8,11,20,22,Gan_Biao_2010,26,27,elfouly2024fitzhugh}.\\ 

\textbf{Noise.} At a macroscopic level noise primarily stems from two sources:
intrinsic thermal fluctuations and extrinsic random disturbances. Additive
noise models the former, capturing the inherent randomness within the system,
while multiplicative noise represents the latter, accounting for variations in
model parameters due to external influences~\cite{6}. \\

Neuronal function, for instance, is inherently noisy, influenced by the
stochastic opening and closing of ion channels, variable presynaptic currents,
and conductivity fluctuations. Lindner \textit{et al.} used the FHN model as prototype
of excitable stochastic dynamics to investigate the effects of Gaussian white
noise on such systems~\cite{92}. Their research spanned isolated units to networks of
coupled elements, unveiling phenomena like noise-induced oscillations,
stochastic resonance, and synchronization, alongside noise-triggered phase
transitions and complex pulse and spiral dynamics, with applications extending
from biophysics to laser technology~\cite{92}.\\

Counterintuitively, random perturbations, especially when combined with weak
deterministic stimuli, can introduce order in temporal and spatial domains of
nonlinear systems, a phenomenon known as stochastic resonance. This effect,
where noise enhances system response to external perturbations, underscores
noise's constructive potential in nonlinear dynamics. Coherence resonance
occurs when an excitable system exhibits maximal signal regularity at optimal
noise levels, even without external driving forces. This phenomenon, alongside
stochastic resonance, has been pivotal in understanding dynamics near
bifurcation points and within bistable and oscillatory systems. Also in the
context of the FHN model, it has sparked significant interest for its
implications in synchronization~\cite{27,104,111,135},
resonance~\cite{18,24,92,94,95,103,111,118,141,155,RevModPhys.79.829}, the
formation and stability of complex spatio-temporal patterns like
spirals~\cite{123,24,172,RevModPhys.79.829,PhysRevLett.91.180601,PhysRevE.65.011105}.\\

\textbf{Transient dynamics.}
Many techniques developed in nonlinear dynamics focus on analyzing the
stability and structure of attracting solutions. However, living systems, in
particular, are most often in
transient~\cite{doi:10.1126/science.aat6412,Rabinovich2008,Verd2014} and
continuously integrate time-varying, noisy information from both internal and
external sources. Currently, there is a lack of a clear framework to
characterize such long transient dynamics of system state trajectories. The
need to adapt the mathematical formalism when describing biological systems in
changing environments has recently been
emphasized~\cite{koch2024ghost,KOCH2024150069,10.1371/journal.pcbi.1011388}.\\

\textbf{Nonreciprocally coupled systems.}
Non-reciprocity is a common feature out of equilibrium, where effective interactions between agents violate Newton's third law. Such interactions are typically observed in living systems, which are often active, with interacting agents converting free energy into directed movement. Despite the widespread importance of non-reciprocal interactions, a general framework for describing their effects is still lacking, hindering our ability to control and exploit them.
In recent years, non-reciprocally coupled systems have received growing
attention~\cite{Fruchart2021,Dinelli2023,PhysRevX.10.041009,PhysRevX.10.041036,PhysRevX.14.021014,PhysRevLett.131.107201}.
Studies on particle and spin models, as well as effective continuum
descriptions, have demonstrated that non-reciprocal systems generically give
rise to spontaneous currents and non-equilibrium patterns that are not
typically seen in the absence of such interactions. These phenomena occur in a
wide range of pattern-forming systems, including mass-conserving
reaction-diffusion systems. Such systems can be mapped to a mass-conserving
FHN model, similar to the non-reciprocal Cahn-Hilliard model, which exhibits
rich spatio-temporal behaviors in oscillatory and excitable
media~\cite{PhysRevX.10.041009,PhysRevX.10.041036,PhysRevX.14.021014,PhysRevLett.131.107201}.\\

\textbf{Multiphysics models.} 
Multiphysics models, integrating various physical processes, are important in simulating complex biological systems, and  cardiac and neuronal systems provide good examples. These models encapsulate interactions across different scales, from molecular to organ levels, offering insights into the complex behaviors of biological systems.\\

\textit{Cardiac models.} In cardiology, multiphysics models blend data and
theories to simulate heart functions, addressing clinical queries with
precision~\cite{31,105,124,130,166}. Balancing model complexity and simplicity
is crucial; overly complex models might not yield more predictive power and
could introduce uncertainties. Models like the FHN system serve as simple yet
powerful tools within these larger frameworks, offering interpretable insights
into cardiac rhythms and anomalies. Advanced applications, such as
fluid-structure interaction simulations for left ventricular dynamics,
leverage the FHN model's simplicity to explore the nuanced interplay between
cellular mechanisms and macroscopic cardiac motions, thus bridging molecular
aberrations with observable clinical disorders~\cite{105}.\\

\textit{Neuronal models.} The brain's complexity requires a multiscale
modeling approach, from single neuron dynamics to network
behaviors~\cite{Deco_2008,Siettos_2016}. The FHN model provides a
straightforward yet potent representation of neuron dynamics and their
integration into larger neuronal models. Bridging the microscopic and
macroscopic scales is key to understanding brain function. Mathematical models,
from biophysical to data-driven, and tools from statistics to dynamical
systems, help decipher the brain's intricate structure and functionality. These
models facilitate hypothesis testing within biological contexts, shedding light
on brain connectivity and dynamics from neurons to networks.\\

\textit{Data-driven modeling.} The intersection of multiphysics models with
rich clinical data promises a new approach to medical diagnostics and
therapeutic strategies, especially in cardiology and neurology. Data-driven
methods are revolutionizing the way models are identified and refined, from
black-box models that offer predictions without transparency to white-box
models that demystify the underlying mechanics through clear mathematical
expressions. Employing techniques like Symbolic Regression~\cite{Schmidt_2009}
or Sparse Identification of Nonlinear Dynamics (SINDy)~\cite{Brunton_2016}, one
can distill simple, yet insightful models characterized by low-order polynomial
differential equations, such as the FHN
model~\cite{PROKOP2024109316,prokop2024datadriven}. Instead of deriving models
as differential equations, there also exist methods to identify regulatory
network interactions from time series. This has for instance also been tested
for the FHN system, revealing the inter- and self-regulations from the
oscillatory changes in its two variables~\cite{tyler2022inferring}.
Such models excel in capturing the essence of complex data, providing a clear
and interpretable framework that can be further refined by incorporating
higher-order corrections when necessary. Hybrid or gray-box methods exploit the
pattern identification strength of black-box models with the interpretability
of white-box approaches by embedding prior knowledge into neural network
designs~\cite{Cranmer_2020,RAISSI2019686,rackauckas2020universal}. The result
is a nuanced equilibrium between empirical data analysis and conceptual rigor,
enriching both the depth and breadth of biological multiphysics modeling.
Anticipating future trends, it is likely that this blend of simple,
interpretable FHN-like models with the computational power of neural networks
will become increasingly prevalent in the exploration of complex biological
systems.

\begin{acknowledgments}
L.G. acknowledges funding by the KU Leuven Research Fund (grant number C14/23/130) and the Research-Foundation Flanders (FWO, grant number G074321N). D.R.-R. is supported by the Ministry of Universities through the “Pla de Recuperacio, Transformació i Resilència” and by the EU (NextGenerationEU), together with the Universitat de les Illes Balears. P.P.-R. is supported by the Sapienza University Grant AddSapiExcellence (NOSTERDIS). We thank Felix Nolet for valuable discussions and input.
\end{acknowledgments}

\section*{Conflict of Interest Statement}

The authors have no conflicts to disclose.\\

\section*{Declaration of generative AI and AI-assisted technologies in the writing process}

During the preparation of this work the author(s) used ChatGPT in order to receive suggestions on grammar and phrasing. All scientific content and figures were created by the authors. After using this tool, the authors reviewed the suggestions and only edited the grammar and phrasing of parts of the text. The authors take full responsibility for the content of the publication.

\section*{Data Availability Statement}

The numerical codes to reproduce the figures in this study are openly available in \textsc{GitLab} \cite{GitLab}, and as an archived repository on RDR by KU Leuven \cite{PF6O4J_2024}.
\nocite{*}

\bibliography{Bibliography}

\end{document}